\documentclass[a4paper,11pt]{article}
\usepackage{amsmath,amssymb,ascmac,enumerate,booktabs,amsthm}
\usepackage{slashed}
\usepackage{braket}
\usepackage{url}
\usepackage{cite}

\def\slashb#1{\not\!\!#1}
\usepackage{hyperref}

\makeatletter
 
  \@addtoreset{equation}{section}
\makeatother

\begin{document}

\begin{titlepage}
\begin{flushright}
TIT/HEP-688 \\
November 2021
\end{flushright}
\vspace{0.5cm}
\begin{center}
{\Large \bf
Global Anomalies
and Bordism Invariants\\
 in One Dimension
}

\lineskip .75em
\vskip 2.5cm
{\large  Saki Koizumi \footnote{sakikoizumi@th.phys.titech.ac.jp}}
\vskip 2.5em
 {\normalsize\it Department of Physics,
Tokyo Institute of Technology,\\
Tokyo, 152-8551, Japan}
\vskip 3.0em


\end{center}


\begin{abstract}
We consider massless Majorana fermion systems with $G=\mathbb{Z}_N$, $SO(N)$, and $O(N)$ symmetry in one-dimensional spacetime.
In these theories, phase ambiguities of the partition functions are given as the exponential of the $\eta$-invariant of the Dirac operators in two dimensions, which is a bordism invariant.
We construct sufficient numbers of bordism invariants to detect all bordism classes.
Then, we classify global anomalies by calculating the $\eta$-invariant of these bordism classes.

\end{abstract}

\end{titlepage}
\baselineskip=0.5cm

\tableofcontents

\section{Introduction}
Gauge anomalies must be canceled  totally for the consistency of a matter theory.
The partition function of a theory defines a homomorphism from an anomalous transformation to an element of $U(1)$.
Then, anomalous gauge transformations are classified as an element of the abelian group.

By the idea of the anomaly inflow \cite{Callan=Harvey, Fad-Shat, Jackiw=CS, Zumino, Stora}, an anomaly of a matter theory in $d$-dimensional spacetime is related to a $(d+1)$-dimensional matter theory.
For example, perturbative fermion anomaly in $d$-dimensional spacetime is related to a Chern-Simons functional on a $(d+1)$-dimensional manifold.
The original $d$-dimensional theory lives on the boundary.
There also exist global gauge anomalies, which were first found in \cite{Old-SU(2)} for a four-dimensional theory with a symmetry of $SU(2)$.
Then, global anomalies in $d$-dimensional spacetime are the Atiyah-Patodi-Singer (APS) $\eta$-invariant \cite{APS1,APS2,APS3} on mapping tori \cite{GG=Witten}.
The mapping torus of a gauge transformation is a $(d+1)$-dimensional manifold determined by the product of the original $d$-dimensional spacetime and $S^1$, where the transition function along $S^1$ direction is the original gauge transformation.

From the Dai-Freed theorem \cite{Dai-Freed,KY,EW}, the phase of the partition function of a massless chiral fermion or Majorana fermions on a $d$-dimensional spacetime $X$ is equivalent to the path integral over a massive fermion system on $(d+1)$-dimensional manifold $Y$ whose boundary theory is the original theory on the $d$-dimensional manifold $X$ \cite{KY=EW}.
In this viewpoint, the partition function of a theory is manifestly gauge-invariant, but the partition function depends on the choice of the bulk spacetime.
Then, global anomalies for a massless chiral or Majorana fermion in $d$ dimensions are the path integrals over the difference of two bulk spacetimes.
In particular, the mapping torus anomalies in \cite{GG=Witten} are interpreted as the difference between two different bulk spacetimes.
This idea can be applied to topological quantum field theories (TQFTs) \cite{Ati-TQFT=88}.
Anomalies in $d$-dimensional TQFT and its relations to $(d+1)$-dimensional symmetry-protected topological (SPT) phases or $(d+1)$-dimensional bordism invariants have been studied in \cite{EW, Ryu-Moo=10, Wen=13, Kapustin2, Kapu-Tho=14, WW=14, Kapustin, CGS=15, Wit=16, JWE-17, Tachikawa=17}.
For each $(d+1)$-dimensional bordism invariant, there exists essentially unique $d$-dimensional TQFT whose partition function is determined by the bordism invariant \cite{Fre-Hop=16, KY=18, FerSPT18}.
The relation between TQFTs and anomalies was studied in \cite{RKY=19, Cor-Ohm=20}.

In this paper, we study global anomalies for massless Majorana fermion.
We first review the anomaly inflow for the massless Dirac fermion system in $d$-dimensional manifold $X$ \cite{KY=EW}.
The partition function is defined as the path integral over the massive Dirac fermion on $(d+1)$-dimensional manifold which satisfy $\partial Y=X$  \cite{KY, EW, KY=EW}:
\begin{align}
S_{d+1}=\int_{Y} d^{d+1}x\sqrt{g}\psi^\dag i(\slashb{D}+m)\psi.
\end{align}
Where the $\slashb{D}:=\gamma^\mu D_\mu$ are the Dirac operator on $Y$, $\gamma^\mu$ is the gamma matrices along the $\mu$ direction, and the boundary condition for the bulk fermion is the Atiyah-Patodi-Singer (APS) boundary condition \cite{KY=EW}:
\begin{align}
L:\:(1-\gamma^{d+1})\psi|_{X}=0.
\end{align}
The path integral over $Y$ is given by
\begin{align}
Z_c(Y)=\int_Y{\cal D}\psi e^{-S_{d+1}}={\rm det}(i\slashb{D}+im)=\prod_j(\lambda_j+im).
\end{align}
Here, we labelled the eigenvalues of ${\cal D}:=i\slashb{D}$ as $\lambda_i$.
In the Euclidian theory with real symmetry group $G$, we cannot determine the complex conjugate of $\psi$.
There exists an antilinear operation ${\rm C}$ on spinor which commutes with the Dirac operator ${\cal D}$ and satisfies ${\rm C}^2=-1$ (See \cite{GG=Witten}, or section $2.2$ and Appendix B in \cite{EW}).
Then, pseudoreal fermions $\psi$ and ${\rm C}\psi$ are independent for any fermions $\psi$.
This causes the doubling of the spectrum of the Dirac operator $(\lambda_i,\lambda_i)$.
In the case $d+1=2$, we choose ${\rm C}$ as ${\rm C}:=\ast\sigma^2$ ($\sigma^2$ is the Pauli matrix).
The case $d+1$ is odd, the chirality operation $\psi\to\gamma^{d+2}\psi$ also gives a paring $(\lambda_i,-\lambda_i)$ for any non-zero eigenvalues.
The partition function of a Majorana fermion on $d$-dimensional spacetime $X$ is given by only the contribution from one of the mode of each pairs $(\lambda_i,\lambda_i)$ \cite{EW, KY=EW}.
If we regularize this path integral by the Pauli-Villars method, the partition function can be written by the $\eta$-invariant of the Dirac operator:
\begin{align}
Z_M(Y)=\prod_j{}'\frac{\lambda_j+im}{\lambda_j-im}=\exp(i\pi\eta)|Z_M|.
\label{ZM}
\end{align}
Where $\prod_j{}'$ is the product of $\lambda_i$ from each pair $(\lambda_i,\lambda_i)$,
and the $\eta$-invariant is determined by the eigenvalues of the Dirac operator $\lambda_i$ \cite{APS1}:
\begin{align}
\eta:=\frac{1}{2}{\rm lim}_{s\to +0}\sum_j{\rm sign}(\lambda_j)e^{-s|\lambda_j|},\qquad
{\rm sign}(\lambda):=\left\{
\begin{array}{ccc}
1&\quad&\lambda\geq 0,\\
-1&&\lambda<0.
\end{array}
\right.
\label{eta}
\end{align}
The partition function eq.(\ref{ZM}) depends on the choice of the bulk spacetime, and therefore the global anomalies are interpreted as the ambiguities of the bulk spacetime choice.
The difference between two partition functions corresponding to two bulk spacetimes $Y_1$ and $Y_2$ is
\begin{align}
\frac{Z_M(Y_1)}{Z_M(Y_2)}
=\frac{\exp(\pi i\eta(Y_1))}{\exp(\pi i\eta(Y_2))}
=\exp(\pi i\eta(Y_1\cup \overline{Y}_2)).
\label{phase}
\end{align}
Here, $\cup$ is the connected sum of the two manifolds, and $\overline{Y}$ is the orientation reversing of a manifold $Y$.
The last equality is satisfied by the Dai-Freed theorem \cite{KY}.
If two bulk spacetimes $Y_1$ and $Y_2$ satisfy $Y_1\cup \overline{Y}_2=\partial Z$, where the spin structures and $G$-bundles over $Y_1$ and $\overline{Y}_2$ are extended to a $(d+2)$-dimensional manifold $Z$, and assume that the perturbative anomalies are cancelled, 
we find that $\exp(\pi i\eta(Y_1\cup \overline{Y}_2))$ is trivial by the APS index theorem \cite{APS1,APS2,APS3}:
\begin{align}
{\rm ind}(i\slashb{D}_Z)=\frac{\eta(Y_1\cup\overline{Y}_2)}{2}.\label{bor=eta1}
\end{align}
Combining eq.(\ref{phase}) and eq.(\ref{bor=eta1}), $\exp(i\pi\eta)$ is a well-defined group homomorphism from $\Omega_{d+1}^{\rm Spin}(BG)$ to $U(1)$:
\begin{align}
\eta[G]:=\exp(\pi i\eta):\:\Omega_{d+1}^{\rm Spin}(BG)\to U(1),\qquad
Y\to \exp(\pi i\eta(Y)).\label{bor-eta-M}
\end{align}
Because all non-zero eigenvalues make quartettes $(\lambda,\lambda,-\lambda,-\lambda)$, only zero-modes contribute to $\exp(i\pi\eta)$:
\begin{align}
\exp(i\pi\eta(Y))=(-1)^{N_Y/2}.
\label{Index-SO(N)}
\end{align}
Here, $N_Y$ is the number of the zero-modes of the Dirac operator on $Y$.
The $\eta$-invariants $\eta[G]$ were studied in many cases.
In the cases $G=\mathbb{Z}_N$ and twisted spin-$\mathbb{Z}_N$ in $d=4$ dimensions were studied in \cite{discrete}.
The Standard Model and the $SU(5)$ and $Spin(10) $ GUTs were considered in \cite{Garcia}.
In those cases, the five-dimensional bordism groups are spanned by five-dimensional lens spaces.
Then, global anomalies are calculated by using the formula of the $\eta$-invariant on lens spaces which are given in \cite{Spherical}, and the Atiyah-Hirzebruch spectral sequence \cite{AHSS}.
The cases two-dimensional bulk spacetimes without additional internal symmetries are considered in \cite{Kapustin, String=WS, Arf-1, Arf-2, Brown=72, ABK-1, ABK-2, Kob=19}.
In the case of two-dimensional spacetime, it was done for some symmetry groups \cite{2d}.

The motivation of this paper is to determine the range of the global anomalies.
In this paper, we will focus on the massless Majorana fermion system with a symmetry $G=\mathbb{Z}_N, SO(N), O(N)$ in one dimension.
In one-dimensional case, eq.(\ref{bor=eta1}) is satisfied, and thus eq.(\ref{bor-eta-M}) is well-defined.
Therefore, to classify global anomalies by the values of global anomalies, we should find all bordism classes, and calculate $\exp(i\pi\eta)$ for each bordism classes.

The rest of this paper is organized as follows.
In section $2$, we will first discuss global anomalies by acting a symmetry transformation on the one-dimensional Hilbert space and explain the relation with global anomalies determined by the mapping tori by the idea of \cite{Old-SU(2)}.
In section $3$, we will construct a bordism invariant which is common for any symmetry group.
In section $4$, section $5$, and section $6$, we will construct some bordism invariants for each symmetry group $G=\mathbb{Z}_n, SO(n), O(n)$, and obtain all bordism classes.
We will also obtain global anomalies for all bordism classes.
In Appendix A, we will show the explicit calculation of the bordism group based on \cite{Garcia} and \cite{DK}.
In Appendix B, we will write down the calculation of the zero-modes on an $O(2)$-spin bundle.
In Appendix C, we will calculate the Stiefel-Whitney class that is a bordism invariant in the case $G=O(N)$.

\section{Old Viewpoint of Global Anomalies}
In this section, we will discuss global anomalies in the old viewpoint.
This section is the review of \cite{KY=EW, Hil=DGG21}.
Let us consider massless Majorana fermion $\chi:=\{\chi^a\}_{a=1,2}$ with a symmetry $G=\mathbb{Z}_N\subset SO(2)$, $SO(N)$, or $O(N)$ ($\chi^a\in\mathbb{R}$ is a symmetry group $G$ component of the Majorana fermion):
\begin{align}
S=\int dx\:\chi_a
i (\partial_x+A(x))^a{}_b \chi^b.
\label{actionL1}
\end{align}
We will first calculate the global anomalies by acting on generators of the gauge symmetries on the Hilbert space, and confirm that these global anomalies are equivalent to the global anomalies determined by the mapping tori in the viewpoint of \cite{Old-SU(2)}.

\subsection{The case $G=SO(2n)$}
We will first determine the global anomalies in the case $G=SO(2n)$.
In the canonical quantization of a fermion $\chi^a$,
we obtain an algebra over $\mathbb{C}$ generated by $\{\chi^a\}_{a=1,\ldots,2n}$ which satsify $\{\chi^a,\chi^b\}=\delta^{ab}$.
If we define
\begin{align}
\psi_\pm^A:=\frac{1}{\sqrt{2}}(\chi^{2A-1}\pm i\chi^{2A}),\qquad
A=1,\ldots,n,\label{psi=pm}
\end{align}
the anti-commutation relation becomes
\begin{align}
\{\psi_+^A,\psi_-^B\}=\delta^{A,B},\qquad
\{\psi_\pm^A,\psi_\pm^B\}=0,\qquad
A,B=1,\cdots,n.
\end{align}
We denote by $\ket{0}$ the vacuum of the Hilbert space that satisfies $\psi_-^A\ket{0}=0$.
 One of the bases of the Hilbert space ${\cal H}$ is as follows:
\begin{align}
{\cal H}:=\{\psi_+^{A_1}\ldots\psi_+^{A_p}\ket{0}\}_{p=0,1,\ldots,n}.\label{basis}
\end{align}
The basis of the dual of the Hilbert space are $\bra{0}\psi_-^{A_1}\ldots\psi_-^{A_p}$, where $p=0,\ldots,n$, and $\bra{0}$ is the dual of the vacuum $\ket{0}$, i.e. $\left<0|0\right>=1$.

Since holonomy of the $SO(2n)$-bundle must be inserted into the partition function, we should first discuss the operation of the $SO(2n)$ symmetry on the Hilbert space.
The $SO(2n)$ generators act on the Hilbert space as
\begin{align}
Q^{ab}:=i\chi^a\chi^b=-Q^{ba},\qquad a\neq b.\label{gen=SO}
\end{align}
We consider the case that the $SO(2n)$ holonomy is given as a maximal torus:
\begin{align}
g(\alpha_1,\ldots,\alpha_{n}):=
\left(
\begin{array}{cccc}
D(\alpha_1)&&\\
&\ddots&\\
&&D(\alpha_{n})
\end{array}
\right)
,\quad
D(\theta):=
\left(
\begin{array}{cc}
\cos\theta&\sin\theta\\
-\sin\theta&\cos\theta
\end{array}
\right).\label{max=SO2n}
\end{align}
A maximal torus $g(\alpha_1,\ldots,\alpha_{n})\in SO(2n)$ act on the Hilbert space as follows:
\begin{align}
g(\alpha_1,\ldots,\alpha_{n})\ket{\psi}
=&\exp(\alpha_1Q_{12})\ldots\exp(\alpha_{n}Q_{N-1,N})\ket{\psi},\qquad
\forall\ket{\psi}\in{\cal H}
\notag\\
=&\prod_A
\exp\left(i\alpha_A(\psi_+^A\psi_-^A-\frac{1}{2})\right)\ket{\psi}
=\exp\left\{i\sum_A\alpha_A\left(\psi_+^A\psi_-^A-\frac{1}{2}\right)\right\}\ket{\psi}.\label{SO(4)rep}
\end{align}
In the same way, we obtain
\begin{align}
\bra{\psi}\left\{g(\alpha_1,\ldots,\alpha_{n})\right\}^\dag=\bra{\psi}\exp\left\{-i\sum_A\alpha_A\left(\psi_+^A\psi_-^A-\frac{1}{2}\right)\right\}.
\label{SO(4)bra}
\end{align}
Therefore, if the spin structure is periodic, the partition function becomes
\begin{align}
Z={\rm Tr}_{\cal H}\;(-1)^Fg(\alpha_1,\ldots,\alpha_{n})
=e^{-\frac{i}{2}\sum_A\alpha_A}
\prod_A(1-e^{i\alpha_A})
=\prod_A\sin(\alpha_A/2).\label{pf2}
\end{align}
If the spin structure is anti-periodic, the partition function is
\begin{align}
Z={\rm Tr}_{\cal H}\;g(\alpha_1,\ldots,\alpha_{n})
=e^{-\frac{i}{2}\sum_A\alpha_A}
\prod_A(1+e^{i\alpha_A})
=\prod_A\cos(\alpha_A/2).\label{pf2-2}
\end{align}

Now we discuss global anomalies.
In the following, we only consider the case that the spin structure is periodic, but we can also discuss in the case that the spin structure is anti-periodic in the same way.
If the spin structure is periodic, a gauge symmetry $h:S^1\to SO(N)$ transform the partition function as
\begin{align}
Z={\rm Tr}_{\cal H}\;(-1)^Fg(\alpha_1,\ldots,\alpha_{n})
\to
{\rm Tr}_{h{\cal H}}\;(-1)^Fg(\alpha_1,\ldots,\alpha_{n}).
\end{align}
Here, we denote by ${\cal H}$ the original basis of the Hilbert space, and also denote by $h{\cal H}$ the basis after transformed by $h:S^1\to  SO(2n)$.
By using eq.(\ref{SO(4)rep}) and eq.(\ref{SO(4)bra}), a symmetry transformation $h=g(\beta_1,\ldots,\beta_n)$ act on states $\psi_+^{A_1}\ldots\psi_+^{A_p}\ket{0,x}$ and $\bra{0,x}\psi_-^{A_1}\ldots\psi_-^{A_p}$ at a time $x\in S^1$ as follows:
\begin{align}
\psi_+^{A_1}\ldots\psi_+^{A_p}\ket{0,x}\to &e^{-\frac{i}{2}\sum_A\beta_A(x)}\prod_{j=1}^pe^{i\beta_{A_j}(x)}\psi_+^{A_1}\ldots\psi_+^{A_p}\ket{0,x},\label{theta=0}
\\
\bra{0,x}\psi_-^{A_1}\ldots\psi_-^{A_p}\to& e^{\frac{i}{2}\sum_A\beta_A(x)}\prod_{j=1}^pe^{-i\beta_{A_j}(x)}
\bra{0,x}\psi_-^{A_1}\ldots\psi_-^{A_p}.
\label{theta=2pi}
\end{align}
We find that anomaly of a symmetry $h=g(\beta_1,\ldots,\beta_n)$ is $(-1)^{\sum_AN_A}$, by substituting $x=0$ and $x=2\pi$ into (\ref{theta=0}) and (\ref{theta=2pi}), where $N_A\in\mathbb{Z}$ is the winding number of $\beta_A$.
By this gauge transformation,  each $\alpha_A$ in eq.(\ref{pf2}) or eq.(\ref{pf2-2}) is shifted as $\alpha_A\to \alpha_A+2\pi N_A$.
This shift does not change the boundary condition of the fermion, but the phase of the partition function may change if $\sum_AN_A$ is odd.
If the spin structure is anti-periodic, this $S^1$ spacetime can be a boundary of some two-dimensional manifold, and thus we can capture this global anomaly by the anomaly inflow.
However, if the spin structure on a $S^1$ is periodic, this $S^1$ cannot be a boundary of any two-dimensional manifolds.
But if we consider two such $S^1$, the combination of these two $S^1$ can be a boundary of a two-dimensional manifold.
Thus, we can discuss this type of global anomaly by the anomaly inflow also in the case that the spin structure on $S^1$ is periodic.

In the case $G=SO(2)$, we can add a counter term, which is given as
\begin{align}
\Delta\Gamma:=\int dx \;a(x),\label{counter}
\end{align}
where the gauge field $A(x)$ is 
\begin{align}
A(x)=a(x)\left(
\begin{array}{cc}
0&-1\\
1&0
\end{array}
\right).
\end{align}
This counter term is changed by a gauge transformation $h=g(\beta(x))$ as $\Delta \Gamma\to \Delta \Gamma+2\pi N_\beta$, where $N_\beta\in\mathbb{Z}$ is the winding number.
Therefore, we can eliminate gauge anomalies by adding this counter term in the case $G=SO(2)$.
Then, if we embed $\mathbb{Z}_N\subset SO(2)$ and consider the gauge symmetry $G=\mathbb{Z}_N$, the $\mathbb{Z}_N$ symmetry also does not have global anomalies up to adding a counter term.

In the case that the spin structure is periodic, we cannot determine the overall sign of non-zero path integral.
In this case, each $\{\chi^a\}_{a=1,\cdots,2n}$ has a zero-mode.
We denote these zero-modes as $\{\chi_0^a\}_{a=1,\cdots,2n}$.
Minimal non-zero path integral is given by inserting all zero-modes:
\begin{align}
{\rm Tr}(-1)^F\chi_0^1\cdots\chi_0^{2n}.\label{Arf-anom}
\end{align}
However, the sign of this path integral depends on the ordering of these zero-modes $\chi_0^k$.
This ambiguity does not relate to global anomalies of symmetry transformations.
But, this type of global anomaly also is interpreted in the viewpoint of the $\eta$-invariant.
If we consider two such $S^1$, the combination of these two $S^1$ can be a boundary of a two-dimensional manifold.
Since each $S^1$ is not a boundary, we cannot determine the sign of ${\rm Tr}(-1)^FU$, where $U$ is the $SO(2n)$ holonomy on the $S^1$.

\subsection{The case $G=SO(2n+1)$}
Next we consider the case $G=SO(2n+1)$.
By the canonical quantization, we obtain an algebra over $\mathbb{C}$ which is generated by $\{\chi^a\}_{a=1,\ldots,2n+1}$ that satisfy $\{\chi^a,\chi^b\}=\delta^{ab}$.
We introduce $\psi_\pm^A$ as eq.(\ref{psi=pm}).
The Hilbert space is constructed by acting $\psi_+^A$'s on the vacuum $\ket{0}$, where the vacuum is eliminated by $\psi_-^A\ket{0}=0$.
Another operator $\chi^{2n+1}$ should act on the vacuum as $\chi_{2n+1}\ket{0}=\pm\ket{0}$.
Here, we denote by $\ket{0}_\pm$ these vacua $\chi_{2n+1}\ket{0}_\pm=\pm\ket{0}_\pm$.
(For example, in the case $n=1$, if we assume $\chi\ket{0}$ is independent of $\ket{0}$, and if we add the same system, the Hilbert space is three dimensions.
This contradicts the fact that the dimension of Hilbert space of $G=SO(2)$ is two.
Thus, we should define $\chi\ket{0}=\pm\ket{0}$).
Therefore, in the case $G=SO(2n+1)$, there are two different Hilbert spaces.

If the spin structure is anti-periodic, the fermion number operator $(-1)^F$ is inserted into the partition function.
Since the fermion number operator exchange these two different Hilbert space by the anti-commuting relation $\{(-1)^F,\chi_{2n+1}\}=0$, the partition function is defined by using both two different Hilbert spaces.
We can define trace in the definition of the partition function by an isomorphism between two different Hilbert spaces, which relate $\psi_+^{A_1}\ldots\psi_+^{A_p}\ket{0}_+$ and $\psi_+^{A_1}\ldots\psi_+^{A_p}\ket{0}_-$.

Now we discuss global anomalies.
In this case, there are global anomalies of the symmetry $G=SO(2n+1)$ as same as global anomalies discussed in the case $G=SO(2n)$.
However, there is another type of anomaly that appears in the case that is related to the existence of odd numbers of zero-modes.
If the spin structure is periodic, non-zero path integral must include all $2n+1$ zero-modes:
\begin{align}
{\rm Tr}(-1)^F\chi_0^1\cdots\chi_0^{2n+1}.\label{Arf2-anom}
\end{align}
Since this path integral includes odd numbers of zero-modes, this path integral changes the sign under the action of $(-1)^F$.
As we explained in section $2.1$, such a $S^1$ cannot be a boundary of any two-dimensional manifold, but if we consider two such a $S^1$, these combination becomes a boundary of a two-dimensional manifold.
Thus, we can treat this global anomaly by the Dai-Freed description.

\subsection{The case $G=O(N)$}
Let us consider the case $G=O(N)$.
 If the transition function of the $O(N)$-bundle valued in $SO(N)$, the partition function is given by eq.(\ref{pf2}) or eq.(\ref{pf2-2}).
 But there is another case that the transition function of the $O(N)$-bundle is as the form $T\;g(\alpha_1,\ldots,\alpha_{[N/2]})$, where $T$ is the projective representation of an element ${\rm diag}(-1,1,\ldots,1)\in O(N)$ on the Hilbert space.
 Because, the $SO(N)\subset O(N)$ part act on the Hilbert space by the spinor representation, we find that $T^2=\pm 1$.
 If we denote by $L_{ab}$ the generator of $SO(N)$ that rotate $(a,b)$-plane, we obtain
\begin{align}
{\rm diag}(-1,1,\ldots,1)L_{1b}
=&-L_{1b}{\rm diag}(-1,1,\ldots,1),
\notag \\
{\rm diag}(-1,1,\ldots,1)L_{ab}
=&L_{ab}{\rm diag}(-1,1,\ldots,1),\qquad
a,b\neq 1.
\end{align}
We determine commutation relations of $T$ and the $SO(N)$ generators $Q^{ab}:=i\chi^a\chi^b=-Q^{ba}$ as 
$\{T,Q_{1,b}\}=0$ and $[T,Q_{ab}]=0$, $a,b\neq 1$, and commute with the Hamiltonian $[T,H]=0$.
Then, $T$ is a linear operator and satisfies $T\chi^a=(-1)^{\delta_{a,1}}\chi^aT$, $a=1,\ldots,[N/2]$, and we find
\begin{align}
T\psi_\pm^1=-\psi_\mp^1T,\qquad
T\psi_\pm^{A\geq 2}=\psi_\pm^{A\geq 2}T.
\end{align}
If the operator $T$ is well-defined on the Hilbert space, a state $T\ket{0}$ is expanded as follows:
\begin{align}
T\ket{0}=\alpha\ket{0}+\alpha_A\psi_+^A\ket{0}+\ldots+\alpha_{1,2,\ldots,n}\psi_+^1\ldots\psi_+^{n}\ket{0}.
\end{align}
If we act $\psi_-^{A\geq 2}$ or $\psi_+^1$ on this expansion, we obtain
\begin{align}
0=&T\psi_-^{A\geq 2}\ket{0}=\psi_-^{A\geq 2}T\ket{0}
=\sum_{\beta_1,\ldots,\beta_j}\alpha_{\beta_1\ldots\beta_j}\psi_-^{A\geq 2}\psi_+^{\beta_1}\ldots\psi_+^{\beta_j}\ket{0},
\notag \\
0=&T\psi_-^1\ket{0}=\psi_+^1T\ket{0}=
\sum_{\beta_1,\ldots,\beta_j}\alpha_{\beta_1\ldots\beta_j}\psi_+^1\psi_+^{\beta_1}\ldots\psi_+^{\beta_j}\ket{0},
\end{align}
and we find
\begin{align}
T\ket{0}=&\beta\psi_+^1\ket{0},\qquad\beta=\pm 1,\pm i,
\notag \\
T\psi_+^{A_1}\ldots\psi_+^{A_p}\ket{0}=&\beta(-1)^{\sum_{j=1}^p\delta_{1,A_j}}\psi_{(-1)^{\delta_{1,A_1}}}^{A_1}\ldots
\psi_{(-1)^{\delta_{1,A_p}}}^{A_p}\psi_+^1\ket{0}.\label{T}
\end{align}
Then, we find $T^2=+1$ if $\beta=\pm 1$, and $T^2=-1$ if $\beta=\pm i$.

There are global anomalies discussed in the case $G=SO(N)$.
There exist other types of global anomalies related to the operation $T$.
The operator $T$ satisfies the following anti-commutation relation:
\begin{align}
\{T,(-1)^F\}=&0.
\end{align}
Since the operator $(-1)^F$ is inserted into the partition function if the spin structure is periodic, $T$ transformation changes the phase of the partition function.
If the spin structure is anti-periodic, the $T$ operation does change the phase of the partition function.

There is also another type of global anomaly.
If we consider a $S^1$ whose spin structure is anti-periodic, and $O(N)$ transition function is given by ${\rm diag}(-1,1,\cdots,1)$, $\chi_1$ has a zero-mode, but $\chi_{k=2,\cdots,N}$ do not have zero-modes.
Thus, we obtain a global anomaly of the symmetry $(-1)^F$ which was already discussed in section $2.2$.
This global anomaly is a mixed anomaly between spin and $O(N)$ symmetry.

\subsection{Relation to Mapping Torus}
We considered global anomalies by the action of symmetries on the Hilbert spaces and confirmed that these global anomalies can be captured by the anomaly inflow based on \cite{KY=EW, Hil=DGG21}.
In the following, we will explain that the global anomalies are given by the $\eta$-invariant on the mapping torus in the case $G=SO(2)$.
We can generalize this result also in the case $G=SO(N)$ where the gauge transformation is valued in the maximal tori.

Let us consider global $SO(2)$ gauge transformation $h(x)=g(\beta(x))$ and the corresponding mapping torus $M_h$.
To obtain the $\eta$-invariant on the mapping torus, we will consider the number of zero-modes on the mapping torus.
The Dirac operator on the mapping torus $M_h$ is given as
\begin{align}
\slashb{D}_{M_h}=\sigma_2\partial_t+\sigma_1(\partial_x+A(x,t)),\qquad
A(x,t)=t\partial_x\beta(x)
\left(
\begin{array}{cc}
0&-1\\
1&0
\end{array}
\right),
\end{align}
where $\sigma_1,\sigma_2$ are the Pauli matrices, which are defined by
\begin{align}
\sigma_1=\left(
\begin{array}{cc}
0&1\\
1&0
\end{array}
\right),\qquad
\sigma_2=\left(
\begin{array}{cc}
0&-i\\
i&0
\end{array}
\right),\qquad
\sigma_3=\left(
\begin{array}{cc}
1&0\\
0&1
\end{array}
\right).
\end{align}
We denote by $\Phi_0(x,t)$ a zero-mode on the mapping torus:
\begin{align}
\partial_t\Phi_0(x,t)=&i\sigma_3(\partial_x+A(x,t))\Phi_0(x,t).\label{1}
\end{align}
At each slice of $t$, we can interpret fermions as fermions on $S^1$.
We denote by $\chi^{j=1,2}(x,t)$ the two $SO(2)$ components of a fermion $\chi(x,t)$ on $S^1$ whose gauge field is given as $A(x,t)$.
These two components satisfy periodic or anti-periodic conditions along the $x$-direction that is determined by the spin structure $\nu=0,1$.
We should also fix the boundary condition along the $t$-direction.
We impose the following boundary condition along the $t$-direction:
\begin{align}
\chi^1(x,1)\pm i\chi^2(x,1)=&e^{\mp i\{\beta(x)-N_\beta x\}}\left\{\chi^1(x,0)\pm i\chi^2(x,0)\right\},
\notag \\
\chi^1(2\pi,t)\pm i\chi^2(2\pi,t)=&(-1)^\nu\left\{\chi^1(0,t)\pm i\chi^2(0,t)\right\}.
\end{align}
Here, $N_\beta$ is the winding number of $\beta(x)$.
By using this notation, $iD_t=i(\partial_x+A(x,t))$ act on the fermion as
\begin{align}
iD_t\chi(x,t)=i\left(
\begin{array}{cc}
\partial_x+it\beta'(x)&0\\
0&\partial_x-it\beta'(x)
\end{array}
\right)
\left(
\begin{array}{c}
\frac{1}{\sqrt{2}}\left\{\chi^1(x,t)+ i\chi^2(x,t)\right\}\\
\frac{1}{\sqrt{2}}\left\{\chi^1(x,t)- i\chi^2(x,t)\right\}
\end{array}
\right).
\end{align}
Then, the eigenvalues of $iD_t$ are labelled as $\lambda_n^\pm(t)=n+\nu/2\pm N_\beta t$, $n\in\mathbb{Z}$, and the corresponding eigenmode $\chi_{n,\pm}(x,t)$ is given as follows:
\begin{align}
\chi_{n,\pm}(x,t)=&\frac{1}{\sqrt{4\pi}}e^{\pm i\{\beta(x)-N_\beta x\}t}e^{-i(n+\nu/2)x}\left(
\begin{array}{c}
1\\
\pm i
\end{array}
\right),
\notag \\
\int_0^{2\pi}dx\chi_{n,\pm}^\dag(x,t)\chi_{m,\pm}(x,t)=&\delta_{n,m},\qquad
\int_0^{2\pi}dx\chi_{n,\pm}^\dag(x,t)\chi_{m,\mp}(x,t)=0.
\end{align}
We also define $\Psi_{n,\pm}(x,t):=(\chi_{n,\pm}(x,t),0)$ and $\tilde{\Psi}_{n,\pm}(x)):=(0,\chi_{n,\pm}(x,t))$. Any zero-modes on the mapping torus are as follows:
\begin{align}
\Phi_0(x,t)=\sum_n \left\{a_n^+(t)\Psi_{n,+}(x,t)+a_n^-(t)\Psi_{n,-}(x,t)+b_n^+(t)\tilde{\Psi}_{n,+}(x,t)+b_n^-(t)\tilde{\Psi}_{n,-}(x,t)\right\}.\label{3}
\end{align}
If we substitute eq.(\ref{3}) into eq.(\ref{1}), we obtain
\begin{align}
\left\{
\partial_t-\left(n+\frac{\nu}{2}\pm N_\beta t\right)\sigma_3\right\}\phi_n^\pm(t)=\pm i\left\{\frac{1}{2\pi}\int dx\beta(x)-\pi N_\beta\right\}\phi^\pm_n(t),\qquad
\phi_n^\pm(t):=(a_n^\pm(t),b_n^\pm(t)).\label{4}
\end{align}
Furthermore, since $\Phi_0(x,1)=h(x)\Phi_0(x,0)$, we find
\begin{align}
a_n^\pm(1)=a_{n\pm N_\beta}^\pm(0),\qquad
b_n^\pm(1)=b_{n\pm N_\beta}^\pm(0).\label{4-2}
\end{align}
Since a fermion is a section of the $SO(2)$ and ${\rm Spin}(2)=U(1)$-bundle, we find that the freedom of the zero-modes is determined by $\{a_n^\pm(t)\}$.
Therefore, by solving eq.(\ref{4}) and eq.(\ref{4-2}), we find that there are $2N_\beta$ zero modes on the mapping torus, and thus we find that $\eta[SO(2)]=(-1)^{N_\beta}$.

We also find that $2N_\beta$ eigenvalues across the zero eigenvalues by the change of $t:0\to 1$, which phenomenon is called the spectral flow.
We can interpret the spectral flow in the viewpoint of the global anomalies studied in the old viewpoint.
Let us assume that the spin structure is periodic. (We can consider the case that the spin structure is anti-periodic in the same way).
By a gauge transformation $h=g(\beta(x))$, we find that each eigenvalues are changed as $n\to n\pm N_\beta$.
Any Majorana fermion on $S^1$ is spanned by 
\begin{align}
\tilde{\chi}_n(x,t):=\chi_{n,+}(x,t)+\chi_{-n,-}(x,t),\qquad n\in\mathbb{Z}.
\end{align}
The basis of the Majorana fermion also changed by $t:0\to 1$ as $\tilde{\chi}_n\to\tilde{\chi}_{n+N_\beta}$.
Then, the order of the modes in the path integral is changed by this transformation.
If we denote by $(n+N_\beta t,n-N_\beta t)$ the eigenvalues of $(\chi_{n,+}(x,t),\chi_{-n,-}(x,t))$, this set is changed as
\begin{align}
(n,-n)\to (n+N_\beta,-n-N_\beta).\label{perm}
\end{align}
Here, we identify $(n,-n)$ and $(-n,n)$.
Thus, the ordering of the modes is changed by this permutation eq.(\ref{perm}).
Since the sign of this cyclic permutation of the fermions is equivalent to $(-1)^{N_\beta}$, we can find that the global anomaly of the gauge transformation $h=g(\beta(x))$ is equivalent to $(-1)^{N_\beta}$ if the spin structure is periodic.
We can also verify that the global anomaly of the gauge transformation $h=g(\beta(x))$ is equivalent to $(-1)^{N_\beta}$ if the spin structure is anti-periodic.
We can also discuss global anomalies of an $SO(N)$ transformation which is valued in the maximal torus in the same way.

In the case $G=SO(2n+1)$, there is also a global anomaly of $(-1)^F$ if the spin structure is periodic (See section $2.2$).
We can construct the mapping torus of a transformation $(-1)^F$ that is determined as the product of $x\in S^1$ and $t\in [0,1]$ interval, where $SO(2n+1)$ partition function on $S^1$ is trivial, and fermions on the mapping torus satisfies $\Psi(x,1)=-\Psi(x,0)$.
 We find that the $\eta$-invariant on the mapping torus is given as $\eta[SO(2n+1)]=-1$.
 This is equivalent to the fact that the number of eigenmodes on $S^1$ that pass through the zero-modes is two.
 Then, we can interpret the spectral flow as the global anomaly of the transformation $(-1)^F$ in the same way as the case $G=SO(2)$.
In the same way, we can verify that the global anomalies of the transformations $(-1)^F$ and $T$ explained in section $2.3$ are given by the $\eta$-invariant on the mapping tori.
However, we can not capture the global anomaly that is not related to a transformation (See section $2.1$) by the mapping torus.

\section{Bordism Invariant: $\Omega_2^{\rm Spin}(BG)\to\Omega_2^{\rm Spin}(pt)\to U(1)$}
In the following, we will study the range of global anomalies in the cases $G=\mathbb{Z}_N,SO(N)$, and $O(N)$ in the description of the Dai-Freed theorem.
For this purpose, we will first construct the sufficient number of bordism invariants in the cases $G=\mathbb{Z}_N,SO(N)$, and $O(N)$.
In this section, we will construct a bordism invariant which is common for any gauge group.

Let us first recall that the forgetful map of a $G$-bundle is well-defined and surjective for all symmetry groups $G$ \cite{DK}:
\begin{align}
\Phi_{pt}:\:\Omega_d^{\rm Spin}(BG)\to\Omega_d^{\rm Spin}(pt).
\label{Phi}
\end{align}
The continuous function from $BG$ to a point $ \pi:BG\to \{pt\}$ introduces a group homomorphism $\pi_\ast:\Omega_d^{\rm Spin}(BG)\to \Omega_d^{\rm Spin}(pt)$.
The inclusion $i:\{pt\}\to BG$ also induce a homomorphism $i_\ast:\Omega_d^{\rm Spin}(pt)\to \Omega_d^{\rm Spin}(BG)$.
From $\pi\circ i=id$, these induced homomorphisms satisfy $\pi_\ast\circ i_\ast=id$, which means $\pi_\ast$ is surjective.
We obtain an exact sequence:
\begin{align}
0\to {\rm ker}\:\pi_\ast\xrightarrow{h}
\Omega_d^{\rm Spin}(BG)
\xrightarrow{\pi_\ast}
\Omega_d^{\rm Spin}(pt)
\to 0.
\end{align}
Then, the bordism group can be divided into 
\begin{align}
\Omega_d^{\rm Spin}(BG)\simeq\Omega_d^{\rm Spin}(pt)\oplus{\rm ker}\:\pi_\ast.
\label{split}
\end{align}
This surjection $\pi_\ast$ is the forgetful map $\Phi_{pt}$ in eq.(\ref{Phi}).
Here, $\tilde{\Omega}_d^{\rm Spin}(BG):={\rm ker}\:\Phi_{pt}={\rm ker}\:\pi_\ast$ is called the reduced bordism group.

We now construct a bordism invariant as a composition of eq.(\ref{Phi}) and a non-trivial bordism invariant of $\Omega_d^{\rm Spin}(pt)$.
In the case $d=2$, we can choose the $\eta$-invariant $\eta[pt]$ in eq.(\ref{bor-eta-M}) as a non-trivial element of the bordism group $\Omega_2^{\rm Spin}(pt)$:
\footnote{Because $\Omega_2^{\rm Spin}(pt)=\mathbb{Z}_2$, there is a unique non-trivial bordism invariant for $d=2$, this $\eta[pt]$ is equivalent to the Arf-invariant \cite{Arf-1},\cite{Arf-2}.}
\begin{align}
\eta[pt]\circ\Phi_{pt}:\:
\Omega_2^{\rm Spin}(BG)
\to
U(1),\qquad
[M(s,g)]\to \exp\left\{i\pi\eta[pt]\circ\Phi_{pt}(M[s,g])\right\}.
\label{eta-ptO(N)}
\end{align}
Here, we denote by $M(s,g)$ a manifold whose spin structure and $G$-bundle correspond to the classifying maps $s$ and $g$ (See section $8.6$ in \cite{DK}), and denote by $[M(s,g)]$ the corresponding element of $\Omega_2^{\rm Spin}(BG)$.
In particular, we will calculate $\eta[pt]$ on a torus $x\sim x+L_x$, $y\sim y+L_y$ with a spin structure.
A fermion on the torus satisfies the following conditions:
\begin{align}
\psi(x+L_x,y)=(-1)^{\nu_1}\psi(x,y)
,\qquad
\psi(x,y+L_y)=(-1)^{\nu_2}\psi(x,y).
\end{align}
Here, the parameters $\nu_1,\nu_2=0,1$ determine the spin structure.
There exist no zero-modes if the spin structure $(\nu_1,\nu_2)$ is not periodic-periodic,
and exists a zero-mode if $(\nu_1,\nu_2)$ is periodic-periodic.
By using the formula (\ref{Index-SO(N)}), $\eta[pt]$ is given as follows:
\begin{align}
\eta[pt]=
\left\{
\begin{array}{ccc}
-1&&(\nu_1,\nu_2)=(0,0),\\
1&\qquad&{\rm otherwise}.
\end{array}
\right.
\label{pf-2}
\end{align}
We obtain $\eta[pt]\circ\Phi_{pt}=+1$ for a torus whose spin structure is not periodic-periodic, 
and $\eta[pt]\circ\Phi_{pt}=-1$ for a torrus whose spin structure is periodic-periodic.

We will construct other bordism invariants to find all bordism classes.
In the following, we will construct a sufficient number of bordism invariants in the cases $G=\mathbb{Z}_n$, $SO(n)$, and $O(n)$.

\section{The case $G=\mathbb{Z}_N$}
We will determine the range of global anomalies in the case $G=\mathbb{Z}_N$.
The bordism groups are given as $\Omega_2^{\rm Spin}(B\mathbb{Z}_{2n+1})=\mathbb{Z}_2$ and $\Omega_2^{\rm Spin}(B\mathbb{Z}_{2n})=\mathbb{Z}_2\oplus\mathbb{Z}_2$, where computation can be found in Appendix A.
To determine the range of global anomalies, we will consider $G=\mathbb{Z}_{2n+1}$ and $G=\mathbb{Z}_{2n}$ separately.

\subsection*{The case $N=2n+1$}
In the case of a massless Weyl fermion with a symmetry group $G=\mathbb{Z}_{2n+1}\subset SO(2)$ in one dimension,
the phase of the partition function is given by eq.(\ref{ZM}).
Global anomalies are given as eq.(\ref{bor-eta-M}).
By using the bordism invariant $\eta[pt]\circ\Phi_{pt}$, a torus with a periodic-periodic spin structure and a torus with a spin structure that is not periodic-periodic are not bordant.
Since $\Omega_2^{\rm Spin}(B\mathbb{Z}_{2n+1})=\mathbb{Z}_2$, we find that tori span all bordism classes.

We will calculate $\eta[\mathbb{Z}_{2n+1}]$, which was defined in eq.(\ref{bor-eta-M}).
The Dirac operator on a torus $x\sim x+L_x$ and $y\sim y+L_y$ with a spin structure and a $\mathbb{Z}_{2n+1}$-bundle is given as 
\begin{align}
i\slashb{D}=i(\sigma^1\partial_1+\sigma^2\partial_2)
=\left(
\begin{array}{cc}
0&i\partial_x+\partial_y\\
i\partial_x-\partial_y&0
\end{array}
\right).
\label{Dirac}
\end{align}
Here, $\sigma_1$ and $\sigma_2$ are the Pauli matrices. 
We embed $\mathbb{Z}_{2n+1}\subset SO(2)$ and write the $SO(2)$ components of the fermion as $\psi_i$, $i=1,2$,
and introduce $\chi=\psi_1+ i\psi_2,\:\tilde{\chi}=\psi_1-i\psi_2$.
The boundary condition for the fermion is given as follows:
\begin{align}
\chi(x+L_x,y)=&(-1)^{\nu_1}e^{- 2\pi iK_1/(2n+1)}\chi(x,y),\qquad
\chi(x,y+L_y)=(-1)^{\nu_2}e^{-2\pi iK_2/(2n+1)}\chi(x,y)
,\notag \\
\tilde{\chi}(x+L_x,y)=&(-1)^{\nu_1}e^{+2\pi iK_1/(2n+1)}\tilde{\chi}(x,y),\qquad
\tilde{\chi}(x,y+L_y)=(-1)^{\nu_2}e^{+2\pi iK_2/(2n+1)}\tilde{\chi}(x,y).\label{K1-K2}
\end{align}
Here, the parameters $K_1,K_2=0,\ldots,2n$ determine the transition functions of the $\mathbb{Z}_{2n+1}$ bundle along two directions on the torus.
$\nu_1,\nu_2=0,1$ determine the spin structure.
There exist four zero-modes in the case $(\nu_1,\nu_2)=(0,0)$ and $K_1=K_2=0$,
and no zero-modes for otherwise.
Then, the $\eta$-invariant eq.(\ref{Index-SO(N)}) on tori becomes trivial:
\footnote{
If the symmetry group is embedded as $G=\mathbb{Z}_N\subset U(1)$, operation ${\rm C}$ explained in the Introduction does not commute with this symmetry $G=\mathbb{Z}_N\subset U(1)$ unless $N=2$.
Then, doubling of the eigenvalues of the Dirac operator does not occur.
The case we embed $G=\mathbb{Z}_N\subset SO(2)$, $\eta[\mathbb{Z}_N]$ is equivalent to the $\eta$-invariant $\exp(2\pi i\eta)$ of a $\mathbb{Z}_N\subset U(1)$ bundle.
}
\begin{align}
\eta[\mathbb{Z}_{2n+1}]=1.
\label{pf-3}
\end{align}
Since tori span all bordism classes, we conclude that and there are no global anomalies in the case $G=\mathbb{Z}_{2n+1}$.

\subsection*{The case $N=2n$}
In the case $G=\mathbb{Z}_{2n}$, the bordism group is given as $\Omega_{2}^{\rm Spin}(B\mathbb{Z}_{2n})=\mathbb{Z}_2\oplus\mathbb{Z}_2$ (See Appendix A).
We have two bordism invariants, $\eta[\mathbb{Z}_{2n}]$ and $\eta[pt]\circ\Phi_{pt}$ defined in eq.(\ref{bor-eta-M}) and eq.(\ref{eta-ptO(N)}).
\footnote{For $G=\mathbb{Z}_2$, we can also consider the Stiefel-Whitney class, where we will introduce in Appendix C. But the Stiefel-Whitney class is trivial for $G=\mathbb{Z}_2$.}
As in the same way as $G=\mathbb{Z}_{2n+1}$ case,
we find that $\eta[\mathbb{Z}_{2n}]$ is trivial on a torus:
\footnote{
We can also consider embedding $\mathbb{Z}_2\subset U(1)$, and consider a Majorana fermion.
Then, the $\eta$-invariant $\exp(i\pi\eta)$ is valued in $\pm 1$, and satisfies $\left(\exp(i\pi\eta)\right)^2=\eta[\mathbb{Z}_2]$.
}
\begin{align}
\eta[\mathbb{Z}_{2n}]=1\quad{\rm on}\:{\rm tori}.\label{pf-4}
\end{align}

We need another bordism invariant to find all bordism classes.
First recall that in the case $N=2$, we can construct a bordism invariant by using the Arf invariant \cite{String=WS}.
In the case $N=2n$, $n\in\mathbb{Z}$, we consider a non-trivial homomorphism $P_n:\mathbb{Z}_{2n}\to\mathbb{Z}_2$.
This homomorphism $P_n$ induce a bundle map (See \cite{DK} p.$220$, Theorem $8.22$):
\begin{align}
\begin{array}{ccc}
E\mathbb{Z}_{2n}&\xrightarrow{EP_n}&E\mathbb{Z}_2\\
p_1\downarrow&&p_2\downarrow\\
B\mathbb{Z}_{2n}&\xrightarrow{BP_n}&B\mathbb{Z}_2
\end{array}
\label{BPn}
\end{align}
Here, the bundle map $EP_n$ satisfies
\begin{align}
EP_n(x\cdot g)=\left(EP_n(x)\right)\cdot P_n(g),\qquad
\forall x\in E\mathbb{Z}_{2n},\qquad g\in \mathbb{Z}_{2n}.
\label{EPn}
\end{align}
$BP_n$ also induces a group homomorphism:
\begin{align}
(BP_n)_\ast:\Omega_2^{\rm Spin}(B\mathbb{Z}_{2n})\to \Omega_2^{\rm Spin}(B\mathbb{Z}_2),\qquad
[M,\phi]\to[M,BP_n\circ\phi].\label{hom-Z2nZ2}
\end{align}
Here, $M$ is a two-dimensional manifold, $\phi:M\to B\mathbb{Z}_{2n}$ is the classification map (See \cite{DK}, p.$220$), and $[M,\phi]$ is the corresponding bordism class.
An element $[M,\phi]$ corresponds to a pull back bundle $\phi^\ast(E\mathbb{Z}_{2n})\to M$, and $[M,BP_n\circ \phi]$ corresponds to a pullback bundle $(BP_n\circ\phi)^\ast(E\mathbb{Z}_2)\to M$.
Then, we obtain a commutivity diagram (See \cite{DK}, p.$93$):
\begin{align}
\begin{array}{ccccccc}
\phi^\ast(E\mathbb{Z}_{2n})&\xrightarrow{\psi_1} &E\mathbb{Z}_{2n}&\xrightarrow{EP_n}&E\mathbb{Z}_2&\xleftarrow{\psi_2}&(BP_n\circ\phi)^\ast(E\mathbb{Z}_2)\\
p'_1\downarrow&&p_1\downarrow&&p_2\downarrow&&p'_2\downarrow\\
M&\xrightarrow{\phi}&B\mathbb{Z}_{2n}&\xrightarrow{BP_n}&B\mathbb{Z}_2&\xleftarrow{BP_n\circ\phi}&M
\end{array}\label{BPn=BZ2}
\end{align}
where $\psi_1$, $\psi_2$ are bundle maps.
By using the commutivity property of the diagram, and the definition of the bundle map, the following linear isomorphisms exist:
\begin{align}
\psi_1|_{p'_1{}^{-1}(x)}:\: p'_1{}^{-1}(x)\xrightarrow{\simeq} (p_1)^{-1}(\phi(x)),\qquad
\psi_2|_{p'_2{}^{-1}(x)}:\:p'_2{}^{-1}(x)\xrightarrow{\simeq} (p_2)^{-1}(BP_n\circ\phi(x)).\label{gauge-gauge}
\end{align}
In diagram eq.(\ref{BPn=BZ2}), the bundle map $\psi_1$ maps a gauge transformation on the pullback bundle $\phi^\ast(E\mathbb{Z}_{2n})\to M$ to a gauge transformation on the bundle 
$p_1:E\mathbb{Z}_{2n}\to B\mathbb{Z}_{2n}$ as the identity.
The same thing holds for the bundle map $\psi_2$.
Therefore, by using eq.(\ref{EPn}) and (\ref{BPn=BZ2}), eq.(\ref{hom-Z2nZ2}) transforms a gauge transformation on $[M,\phi]$ to a gauge transformation on $[M,BP_n\circ \phi]$ as the map $P_n$.

We construct a bordism invariant as the composition of eq.(\ref{hom-Z2nZ2}) and a bordism invariant of $\Omega_2^{\rm Spin}(B\mathbb{Z}_2)$:
\begin{align}
{\rm Arf}[\mathbb{Z}_2]\circ (BP_n)_\ast:\Omega_2^{\rm Spin}(B\mathbb{Z}_{2n})\to \Omega_2^{\rm Spin}(B\mathbb{Z}_2)\to U(1).\label{Z2=det=m}
\end{align}
Here, ${\rm Arf}[\mathbb{Z}_2]:\Omega_2^{\rm Spin}(B\mathbb{Z}_2)\to U(1)$ is a bordism invariant which is determined by using the Arf-invariant (See section $2.1.2$ in \cite{String=WS}):
\footnote{
In the case $N=2$, we can consider embedding $\mathbb{Z}_2\subset U(1)$ and consider a Majorana fermion.
Then, the $\eta$-invariant $\exp(i\pi\eta)$ is equivalent to ${\rm Arf}[\mathbb{Z}_2]$.}
\begin{align}
{\rm Arf}[\mathbb{Z}_2]:\Omega_2^{\rm Spin}(B\mathbb{Z}_2)\to U(1),\qquad
[M(s,g)]\to(-1)^{{\rm Arf}(M,\sigma_s+\sigma_g)}.
\end{align}
Here, we denote by $M(s,g)$ a manifold whose spin structure and $\mathbb{Z}_2$-bundle are determined by classifying maps $s$ and $g$, and use the fact that these classifying maps determine elements $\sigma_s,\sigma_g\in H^1(M,\mathbb{Z}_2)$ (background gauge fields).
In particular, if we calculate two bordism invariants $\eta[pt]\circ\Phi_{pt}$ and ${\rm Arf}[\mathbb{Z}_2]\circ (BP_n)_\ast$ on tori, we find that all bordism classes of $\Omega_{2}^{\rm Spin}(B\mathbb{Z}_{2n})=\mathbb{Z}_2\oplus\mathbb{Z}_2$ are spanned by tori as listed in Table \ref{BZ2n-rep}.
\begin{table}[h]
  \caption{All bordism classes in $\Omega_2^{\rm Spin}(B\mathbb{Z}_{2n})$ spanned by tori}
  \label{BZ2n-rep}
  \centering
  \begin{tabular}{|c|c|}
     \hline
     Tori&$(\eta[pt]\circ\Phi_{pt},{\rm Arf}[\mathbb{Z}_2]\circ (BP_n)_\ast)$\\
     \hline\hline
     $(\nu_1,\nu_2)=(0,0)$ and $(K_1+\nu_1,K_2+\nu_2)=(0,0)$ mod $2$ & $(\eta[pt]\circ\Phi_{pt},{\rm Arf}[\mathbb{Z}_2]\circ (BP_n)_\ast)=(-1,-1)$\\
     \hline
     $(\nu_1,\nu_2)=(0,0)$ and $(K_1+\nu_1,K_2+\nu_2)\neq(0,0)$ mod $2$ & $(\eta[pt]\circ\Phi_{pt},{\rm Arf}[\mathbb{Z}_2]\circ (BP_n)_\ast)=(-1,+1)$\\
      \hline
       $(\nu_1,\nu_2)\neq(0,0)$ and $(K_1+\nu_1,K_2+\nu_2)=(0,0)$ mod $2$ & $(\eta[pt]\circ\Phi_{pt},{\rm Arf}[\mathbb{Z}_2]\circ (BP_n)_\ast)=(+1,-1)$\\
       \hline
       $(\nu_1,\nu_2)\neq(0,0)$ and $(K_1+\nu_1,K_2+\nu_2)\neq(0,0)$ mod $2$ & $(\eta[pt]\circ\Phi_{pt},{\rm Arf}[\mathbb{Z}_2]\circ (BP_n)_\ast)=(+1,+1)$
       \\\hline
  \end{tabular}
\end{table}
Here, $K_1,K_2=0,\ldots,2n-1$ determine the $\mathbb{Z}_{2n}$ transition functions in the same way as eq.(\ref{K1-K2}).

\section{The case $G=SO(N)$}
In the case $G=SO(N)$, the bordism groups are $\Omega_2^{\rm Spin}(BSO(N\geq 3))=\mathbb{Z}_2\oplus\mathbb{Z}_2$ and $\Omega_2^{\rm Spin}(BSO(2))=\mathbb{Z}_2\oplus\mathbb{Z}$ as we will calculate in the Appendix A.
We first discuss the case $G=SO(N\geq 3)$, and secondly we consider the case $G=SO(2)$.

\subsection*{The case $N\geq 3$}
In the case $G=SO(N\geq 3)$, we found two bordism invariants, $\eta[pt]\circ\Phi_{pt}$ and $\eta[SO(N\geq 3)]$ defined in eq.(\ref{pf-2}) and eq.(\ref{bor-eta-M}), and the second Stiefel-Whitney class which will be defined in Appendix C.
The value of $\eta[pt]\circ\Phi_{pt}$ on a torus was given in section $3$.
 We will consider $\eta[SO(N\geq 3)]$ and the second Stiefel-Whitney classes on tori.

Let us calculate the $\eta[SO(N\geq 3)]$.
A fermion on a torus $x\sim x+L_x$, $y\sim y+L_y$ satisfies the following conditions:
\begin{align}
\psi(L_x,y)=(-1)^{\nu_x}g_1(y)\psi(0,y),\qquad
\psi(x,L_y)=(-1)^{\nu_y}g_2(x)\psi(x,0).\label{cover=SO}
\end{align}
Here, $\nu_x,\nu_y=0,1$ determine a spin structure, and
$g_1,g_2:S^1\to SO(N)$ are the transition functions of an $SO(N)$-bundle on the torus.
We consider the case that $g_1,g_2\in SO(N)$ are in the element of the maximal torus:
\begin{align}
g_1(y)=g(\alpha_1(y) ,\ldots,\alpha_{[N/2]}(y)),\qquad
g_2(x)=g(\beta_1(x),\ldots,\beta_{[N/2]}(x) ),
\label{pq}
\end{align}
where $\alpha_j(y)$ and $\beta_j(x)$ are periodic along the $S^1$ direction.
The case $N=2n$, $g(\theta_1,\ldots,\theta_n)$ is defined as (\ref{max=SO2n}).
The case $N=2n+1$, we define
\begin{align}
g(\theta_1,\ldots,\theta_{n}):=
\left(
\begin{array}{ccccc}
D(\theta_1)&&&\\
&\ddots&&\\
&&D(\theta_{n})&\\
&&&1
\end{array}
\right).
\end{align}
We introduce the $SO(N)$ components of a fermion as $\psi=(\chi_i)_{i=1,\ldots,N}$, and define $\{\psi_j,\tilde{\psi}_j\}_{j=1,\ldots,[N/2]=n}$ as
\begin{align}
\psi_j:=\chi_{2j-1}+i\chi_{2j}
,\qquad
\tilde{\psi}_j:=\chi_{2j-1}-i\chi_{2j},\qquad
j=1,\ldots,[N/2]=n.
\end{align}
Then, the zero-mode equation becomes $N$ $U(1)$ zero-mode equations of $\{\psi_j,\tilde{\psi}_j\}_{j=1,\ldots,[N/2]=n}$, and the field strength becomes diagonal, i.e. $F={\rm diag}(F_1,\tilde{F}_1,\ldots,F_{[N/2]},\tilde{F}_{[N/2]})$, which satisfies
\begin{align}
\int_{T^2}F_j=-\int_{T^2}\tilde{F}_j=2\pi (p_j-q_j)\in 2\pi\mathbb{Z}.
\label{int-F}
\end{align}
Here, $p_j\in\mathbb{Z}$ and $q_j\in\mathbb{Z}$ are the winding numbers of $\alpha_j$ and $\beta_j$ along the $S^1$ direction.
We denote by $n_\pm(j)$ and $\tilde{\psi}_j$ the dimension of positive/negative chirality zero-mode space for $\psi_j$ and $\tilde{\psi}_j$.
By the Atiyah-Singer index theorem \cite{AS=index}, $n_\pm(j)$ and $\tilde{n}_\pm(j)$ satisfy 
\begin{align}
n_+(j)-n_-(j)=-\tilde{n}_+(j)+\tilde{n}_-(j)=p_j-q_j.
\end{align}
By the bijection $\psi\to {\rm C}\psi$ \cite{KY=EW}, we find $n_\pm(j)=\tilde{n}_\mp(j)$.
Here, ${\rm C}$ is the charge conjugation determined as ${\rm C}:=\ast\sigma^2$, and $\ast$ is the complex conjugation.
Then, the number of zero-modes on the torus is
$\sum_{j=1}^{[N/2]}2(p_j-q_j)$ ${\rm mod}\:4$.
Therefore, eq.(\ref{Index-SO(N)}) becomes
\begin{align}
\eta[SO(N)]=(-1)^{\sum_j(p_j-q_j)}.
\label{T2SO(N)}
\end{align}

To summarize, tori span all bordism classes $\Omega_2^{\rm Spin}(BSO(N\geq 3))=\mathbb{Z}_2\oplus\mathbb{Z}_2$ as Table \ref{BSO(N)-rep}.
This result gives global anomalies for mapping tori whose transition functions are given by eq.(\ref{pq}).
\begin{table}[h]
 \caption{All bordism classes of $\Omega_2^{\rm Spin}(BSO(N\geq 3))$ are spanned by tori}
  \label{BSO(N)-rep}
  \centering
  \begin{tabular}{|c|c|}
     \hline
     Tori&$(\eta[pt]\circ\Phi_{pt},\eta[SO(N\geq 3)])$\\
     \hline\hline
     $(\nu_1,\nu_2)=(0,0)$ and $\sum_j(p_j-q_j)=1$ mod $2$ & $(\eta[pt]\circ\Phi_{pt},\eta[SO(N)])=(-1,-1)$\\
     \hline
      $(\nu_1,\nu_2)=(0,0)$ and $\sum_j(p_j-q_j)=0$ mod $2$ & $(\eta[pt]\circ\Phi_{pt},\eta[SO(N)])=(-1,+1)$\\
      \hline
       $(\nu_1,\nu_2)\neq(0,0)$ and $\sum_j(p_j-q_j)=1$ mod $2$ & $(\eta[pt]\circ\Phi_{pt},\eta[SO(N)])=(+1,-1)$\\
       \hline
       $(\nu_1,\nu_2)\neq(0,0)$ and $\sum_j(p_j-q_j)=0$ mod $2$ & $(\eta[pt]\circ\Phi_{pt},\eta[SO(N)])=(+1,+1)$
       \\\hline
  \end{tabular}
\end{table}

As we will explain in Appendix C, the Kronecker pairing of the second Stiefel-Whitney class and the fundamental class is another type of bordism invariant.
In the following, we calculate the second Stiefel-Whitney class $w_2(E)$ of an $SO(N)$-bundle $E\to T^2$ by using the transition functions.
Let us take a good cover ${\cal U}$ over a manifold $X$.
Then, the following isomorphism is satisfied:
\begin{align}
H^p(X,\underline{O(N)})\simeq H^p({\cal U},\underline{O(N)}).\label{check1}
\end{align}
In the case $X=S^1$, $x\sim x+2\pi$, we can construct a good cover $\{U_j\}_{j=1,2,3}$  over $S^1$ as $U_1=\{x\in S^1|x\in(0-\epsilon,\frac{2\pi}{3}+\epsilon)\}$, $U_2=\{x\in S^1|x\in(\frac{2\pi}{3}-\epsilon,\frac{4\pi}{3}+\epsilon)\}$, $U_3=\{x\in S^1|x\in(0-\frac{4\pi}{3},2\pi+\epsilon)\}$, where $\epsilon>0$ is some small real number.
If we denote by $\{U_j\}_{j=1,2,3}$ and $\{V_j\}_{j=1,2,3}$ these good covers over two $S^1$, we can construct a good cover ${\cal U}$ over a torus $T^2$ as
\begin{align}
{\cal U}:=\{U_{ij}\}_{i,j=1,2,3},\qquad
U_{ij}:=U_i\times V_j,\qquad i,j=1,2,3.\label{good=co}
\end{align}
We can treat the above transition functions $g_1$ and $g_2$ in eq.(\ref{cover=SO}) and eq.(\ref{pq}) as transition functions over this good cover eq.(\ref{good=co}).
For example, we give the transition function from $U_{11}$ to $U_{33}$ as $g_2g_1$.
By the isomorphism eq.(\ref{check1}), $H^1(T^2,\underline{O(N)})$ classify $O(N)$-bundles on a torus with this good cover ${\cal U}=\{U_{ij}\}_{i,j=1,2,3}$.
Furthermore, an exact sequence $1\to\mathbb{Z}_2\to{\rm Pin}_+(N)\xrightarrow{{\rm Ad}}O(N)\to 1$ induce the following exact sequence:
\begin{align}
\cdots\to H^1(X,\underline{{\rm Pin}_+(N)})\to H^1(X,\underline{O(N)})\xrightarrow{\delta^\ast}H^2(X,\mathbb{Z}_2)\to\cdots.
\end{align}
We denote by $g_E\in H^1(X,\underline{O(N)})$ the cohomology class determined by the transition function of an $O(N)$-bundle $E\to X$ with a good cover.
Then, we define
\begin{align}
\omega_2(E):=\delta^\ast(g_E)\in H^2(X,\mathbb{Z}_2).
\end{align}
By the definition, $\omega_2(E)$ is trivial if and only if this $O(N)$-bundle has a ${\rm Pin}_+$ structure.
Because the condition that the second Stiefel-Whitney class $w_2(E)$ is trivial is equivalent to the condition that the $O(N)$-bundle is trivial, we find $w_2(E)=\omega_2(E)$.

Now, we calculate the second Stiefel-Whitney class.
We consider an $SO(N)$-bundle whose transition functions are given by eq.(\ref{pq}).
The transition functions $g_1,g_2$ are lifted to ${\rm Spin}(N)$-valued transition functions by the adjoint map:
\begin{align}
\tilde{g}_x=&\exp\left(\frac{i\alpha_1(y)}{4}[\gamma_1,\gamma_2]+\cdots+\frac{i\alpha_{[N/2]}(y)}{4}[\gamma_{2[N/2]-1},\gamma_{2[N/2]}]\right),
\notag \\
\tilde{g}_y=&\exp\left(\frac{i\beta_1(x)}{4}[\gamma_1,\gamma_2]+\cdots+\frac{i\beta_{[N/2]}(x)}{4}[\gamma_{2[N/2]-1},\gamma_{2[N/2]}]\right),
\end{align}
where $\{\gamma_j\}_{j=1,\ldots,2[N/2]}$ are the gamma matrices which satisfy $\{\gamma_j,\gamma_k\}=2\delta_{jk}$.
These transition functions satisfy the cocycle condition if $\sum_i(p_i-q_i)=0$ mod $2$, and do not satisfy the cocycle condition if $\sum_i(p_i-q_i)=1$ mod $2$.
Therefore, tori which satisfies $\sum_i(p_i-q_i)=0$ mod $2$ are trivial element of $H^2(T^2,\mathbb{Z}_2)=\mathbb{Z}_2$, and tori with $\sum_i(p_i-q_i)=1$ mod $2$ belong to the generator of $H^2(T^2,\mathbb{Z}_2)=\mathbb{Z}_2$.
We will see in section $6.2$ that the Kronecker pairing of the second Stiefel-Whitney class and the fundamental class on these tori are equivalent to the value $H^2(T^2,\mathbb{Z}_2)=\mathbb{Z}_2$.
By comparing the result summarized in the table \ref{BSO(N)-rep}, the Kronecker pairing of the second Stiefel-Whitney class and the fundamental class is equivalent to the $\eta[SO(N)]$ as a bordism invariant.

\subsection*{The case $N=2$}
Let us next consider the case $N=2$.
We found two bordism invariant $\eta[pt]\circ\Phi_{pt}$ and $\eta[SO(2)]$, and calculated these invariants on tori in eq.(\ref{pf-2}) and eq.(\ref{T2SO(N)}).
Because $\Omega_2^{\rm Spin}(BU(1))=\mathbb{Z}_2\oplus\mathbb{Z}$, we need to construct another bordism invariants to find all bordism classes.

Any complex line bundles on a topological space $X$ are classified by the first Chern-class $c_1(L)\in H^2(X;\mathbb{Z})$.
If two two-dimensional spin manifolds $X_1$ and $X_2$ with $U(1)$-bundles are bordant, there exist a three dimensional spin manifold $Z$ that satisfies
$X_1\cup\overline{X}_{2}=\partial Z$, and $U(1)$-bundle and spin structures on $X_1$ and $\overline{X}_2$ extended to $Z$.
Here, we denote by $\overline{X}_{2}$ the orient reversing of $X_2$, and $\cup$ as the connected sum.
By using Stoke's theorem, and using the fact that the $U(1)$ field-strength $F$ and gauge field $A$ satisfy $F=dA$, the first Chern-class of these two manifolds $X_1$ and $X_2$ are equivalent.
Thus, we construct a bordism invariant as follows:
\begin{align}
c_1[U(1)]:\:\Omega_2^{\rm Spin}(BU(1))\to
\mathbb{R};
\qquad
[X,\phi_s,f_{U(1)}]\to
c_1(X,f_{U(1)})=\frac{1}{2\pi}\int_{X} F\in H^2(X;\mathbb{Z}).
\label{BU(1)-c1}
\end{align}
Here, we denote by $X$ a two-dimensional manifold, $\phi_s$ is the classifying map of a spin structure on $X$, $f_{U(1)}$ is the classifying map of a $U(1)$-bundle over $X$, and $[X,\phi_s,f_{U(1)}]\in\Omega_2^{\rm Spin}(BU(1))$ is the corresponding bordism class.
We find that tori span the bordism group $\Omega_2^{\rm Spin}(BU(1))=\mathbb{Z}_2\oplus\mathbb{Z}$ as Table \ref{BU(1)-rep}.
By eq.(\ref{T2SO(N)}), we also find that $\eta[SO(2)]=c_1[U(1)]$ mod $2$.
As mentioned in section $2.1$, this phase ambiguity $\eta[SO(2)]$ of the partition function is canceled by adding a counter term eq.(\ref{counter}), and thus there are no global anomalies in the case $G=SO(2)$.
\begin{table}[h]
 \caption{All bordism classes of $\Omega_2^{\rm Spin}(BU(1))$ are spanned by tori}
  \label{BU(1)-rep}
  \centering
  \begin{tabular}{|c|c|}
     \hline
     Tori&$(\eta[pt]\circ\Phi_{pt},c_1[U(1)])$\\
     \hline\hline
          $(\nu_1,\nu_2)=(0,0)$ and $k\in H^2(T^2,\mathbb{Z})=\mathbb{Z}$ & $(\eta[pt]\circ\Phi_{pt},c_1[U(1)])=(-1,k)$\\
      \hline
             $(\nu_1,\nu_2)\neq(0,0)$ and $k\in H^2(T^2,\mathbb{Z})=\mathbb{Z}$ & $(\eta[pt]\circ\Phi_{pt},c_1[U(1)])=(+1,k)$
       \\\hline
  \end{tabular}
\end{table}

\section{The case $G=O(N)$}
We finally consider the case $G=O(N)$.
We will construct a new bordism invariant that relates to $\eta[\mathbb{Z}_2]$ and calculate the Second Stiefel-Whitney class.
Then, we will show that tori span the bordism group.
After that, we will calculate the $\eta$-invariant $\eta[O(N)]$ on tori, and determine the range of the global anomalies.

\subsection{Bordism Invariant: $\Omega_2^{\rm Spin}(BO(N))\to\Omega_2^{\rm Spin}(B\mathbb{Z}_2)\to U(1)$}
We construct a new bordism invariant which is related to a bordism invariant of $\Omega_2^{\rm Spin}(B\mathbb{Z}_2)$.
We introduce a group homomorphism:
\begin{align}
{\rm det}:\:O(N)\to \mathbb{Z}_2=\{\pm 1\},\qquad
X\to {\rm det}X.\label{det}
\end{align}
In the same way as eq.(\ref{hom-Z2nZ2}), this homomorphism eq.(\ref{det}) induce a group homomorphism:
\begin{align}
B{\rm det}_\ast:\Omega_2^{\rm Spin}(BO(N))\to \Omega_2^{\rm Spin}(B\mathbb{Z}_2),\qquad
[M,\phi]\to[M,B{\rm det}\circ\phi],\label{hom-O(N)Z2}
\end{align}
where $M$ is a two-dimensional spin manifold, $\phi:M\to BO(N)$ is the classifying map, $B{\rm det}:BO(N)\to B\mathbb{Z}_2$ is a bundle map induced by homomorphism eq.(\ref{det}) (See \cite{DK} p.$220$, Theorem $8.22$), and $[M,\phi]$ and $[M,B{\rm det}\circ\phi]$ are the corresponding bordism classes.
As in the same way in the case $G=\mathbb{Z}_{2n}$ explained in section $3$,
(\ref{hom-O(N)Z2}) transforms a gauge transformation on $[M,\phi]$ to a gauge transformation of $[M,B{\rm det}\circ \phi]$ by the map ${\rm det}$.
We construct a bordism invariant as the composition of eq.(\ref{hom-O(N)Z2}) and ${\rm Arf}[\mathbb{Z}_2]$:
\begin{align}
{\rm Arf}[\mathbb{Z}_2]\circ B{\rm det}_\ast:\Omega_2^{\rm Spin}(BO(N))\to& \Omega_2^{\rm Spin}(B\mathbb{Z}_2)\to U(1).\label{Z2=det=m}
\end{align}

We will now show $\Omega_2^{\rm Spin}(BO(N))=3\mathbb{Z}_2$.
We introduce a group homomorphism:
\begin{align}
\tau:\mathbb{Z}_2\to O(N),\qquad
\pm 1\to {\rm diag}(\pm 1,1,\ldots,1).\label{tau}
\end{align}
This map satisfies ${\rm det}\circ \tau=id$.
 We obtain a commuting diagram of bundle maps (See \cite{DK}, p.$220$).
\begin{align}
\begin{array}{ccccc}
E\mathbb{Z}_2&\xrightarrow{E\tau}&EO(N)&\xrightarrow{E{\rm det}}&E\mathbb{Z}_2\\
p_2\downarrow&&p_1\downarrow&&p_2\downarrow\\
B\mathbb{Z}_2&\xrightarrow{B\tau}&BO(N)&\xrightarrow{B{\rm det}}&B\mathbb{Z}_2
\end{array}
\label{Bdet2}
\end{align}
Here, $E{\rm det}$ and $E\tau$ satisfies
\begin{align}
E{\rm det}(x\cdot g)=&\left(E{\rm det}(x)\right)\cdot {\rm det}(g),\qquad
\forall x\in EO(N),\qquad g\in O(N),
\notag\\
E\tau(\tilde{x}\cdot \tilde{g})=&\left(E\tau(\tilde{x})\right)\cdot\tau(\tilde{g}),\qquad
\forall \tilde{x}\in E\mathbb{Z}_2,\qquad\tilde{g}\in\mathbb{Z}_2.\label{Etau=O(N)}
\end{align}
By using ${\rm det}\circ \tau=id$, the composition $B{\rm det}\circ B\tau:B\mathbb{Z}_2\to B\mathbb{Z}_2$ is homotopic to identity (See \cite{DK}, p.$152$).
Then, $B{\rm det}_\ast$ and $B\tau_\ast$ satisfies $B{\rm det}_\ast\circ B\tau_\ast=id$,
where $B\tau_\ast:\Omega_2(B\mathbb{Z}_2)\to\Omega_2^{\rm Spin}(BO(N))$ is a homomorphism induced by $B\tau$.
Thus, $B{\rm det}_\ast$ is surjective, and obtain a short exact sequence:
\begin{align}
0\to{\rm ker}\:(B{\rm det}_\ast)\xrightarrow{f}\Omega_2^{\rm Spin}(BO(N))\xrightarrow{B{\rm det}_\ast}
\Omega_2^{\rm Spin}(B\mathbb{Z}_2)\to 0.
\end{align}
Since $B\tau_\ast$ is injective, we obtain
\begin{align}
\Omega_2^{\rm Spin}(BO(N))\xrightarrow{\simeq}& \Omega_2^{\rm Spin}(B\mathbb{Z}_2)\oplus{\rm Ker}(B{\rm det}_\ast),
\notag\\
[M,\phi]\to&
\left(
 [M,B\tau\circ B{\rm det}\circ\phi]
,\:[M,\phi]-B\tau_\ast[M,B{\rm det}\circ\phi]
\right).
\label{BZ2+Ker}
\end{align}
$B\tau_\ast$ transforms a $\mathbb{Z}_2$ gauge transformations to an $O(N)$ gauge transformation as $\mathbb{Z}_2\xrightarrow{\tau} O(N)$.

By the AHSS, the bordism group is $\Omega_2^{\rm Spin}(BO(N))=3\mathbb{Z}_2$ or $\Omega_2^{\rm Spin}(BO(N))=\mathbb{Z}_2\oplus \mathbb{Z}_4$, which are shown in Appendix A.
Combining this fact with eq.(\ref{BZ2+Ker}) and $\Omega_2^{\rm Spin}(B\mathbb{Z}_2)=\mathbb{Z}_2\oplus\mathbb{Z}_2$, we find that $\Omega_2^{\rm Spin}(BO(N))=3\mathbb{Z}_2$.
By the calculation of AHSS in Appendix A, bordism group $\Omega_2^{\rm Spin}(BO(N))$ can be written as the $E^2$ page of the AHSS as follows:
\begin{align}
\Omega_2^{\rm Spin}(BO(N))
=&E^2_{2,0}\oplus E^2_{1,1}\oplus E^2_{0,2}
=3\mathbb{Z}_2
,\qquad
E^2_{p,q}=H_p(BO(N),\Omega_q^{\rm Spin}(pt)).
\label{AHSS=direct}
\end{align}
We obtain the following isomorphisms:
\begin{align}
E^2_{0,2}=\Omega_2^{\rm Spin}(pt),\qquad
E^2_{1,1}=\tilde{\Omega}_2^{\rm Spin}(B\mathbb{Z}_2),\qquad
E^2_{2,0}={\rm Ker}(B{\rm det}_\ast).
\label{AHSS=2}
\end{align}
\begin{proof}[Proof of eq.(\ref{AHSS=2})]
By Appendix A, $\Omega_2^{\rm Spin}(B\mathbb{Z}_2)$ divides into the direct sum:
\begin{align}
\Omega_2^{\rm Spin}(B\mathbb{Z}_2)=\tilde{E}^2_{2,0}\oplus\tilde{E}^2_{1,1}\oplus\tilde{E}^2_{0,2},\qquad
\tilde{E}^2_{0,2}=\tilde{E}^2_{1,1}=\mathbb{Z}_2
,\qquad
\tilde{E}^2_{2,0}=0
.\label{AHSSBZ2}
\end{align}
Here, we denote by $\tilde{E}^2_{p,q}$ the $E^2$ page of the AHSS of $\Omega_d^{\rm Spin}(B\mathbb{Z}_2)$.
These $E^2$ pages satisfy $\Omega_2^{\rm Spin}(pt)=\tilde{E}^2_{0,2}=\mathbb{Z}_2$ and $\tilde{\Omega}_2^{\rm Spin}(B\mathbb{Z}_2)=\tilde{E}^2_{1,1}=\mathbb{Z}_2$.
$B{\rm det}_\ast$ act on these $E^2$ pages as follows:
\begin{align}
B{\rm det}_\ast:\:E^2_{p,q}=H_p(BO(N),\Omega_q^{\rm Spin}(pt))\to \tilde{E}^2_{p,q}=H_p(B\mathbb{Z}_2,\Omega_q^{\rm Spin}(pt)).\label{E2=tE2}
\end{align}
Since the map $B{\rm det}_\ast$ is surjective,
$B{\rm det}_\ast$ is isomorphism for $(p,q)=(0,2),(1,1)$, and zero for $(p,q)=(2,0)$.
\end{proof}

\subsection{Second Stiefel-Whitney class}
We can also calculate the second Stiefel-Whitney class.
In section $5$, we calculated the second Stiefel-Whitney class on tori whose transition functions along the two directions $g_1$ and $g_2$ are valued in $SO(N)$.
We will discuss the second Stiefel-Whitney class on a torus whose transition functions are not valued in $SO(N)$.
Let us first consider the case $N=2$ on a torus.
Any element of the $O(2)$ group can be written as 
\begin{align}
g_\pm(\theta):={\rm diag}(\pm 1,1)g(\theta),\qquad
g(\theta):=\left(
\begin{array}{cc}
\cos\theta&\sin\theta\\
-\sin\theta&\cos\theta
\end{array}
\right).\label{gpm}
\end{align}
We assume that the $O(2)$ transition functions along the two directions $x$ and $y$ are written as
\begin{align}
g_1(y)=g_\rho(2\pi N_\alpha y/L_y),
\qquad
g_2(x)=g_\kappa(2\pi N_\beta x/L_x),\qquad
\rho,\kappa=\pm,\qquad
N_\alpha,N_\beta\in\mathbb{Z}.\label{gpm=2}
\end{align}
Here, we choose the good cover on a torus introduced in eq.(\ref{good=co}).
A fermion satisfies
\begin{align}
\psi(L_x,y)=(-1)^{\nu_1}g_1(y)\psi(0,y),\qquad
\psi(x,L_y)=(-1)^{\nu_2}g_2(x)\psi(x,0).
\label{tf-1}
\end{align}
Here, the parameters $\nu_1,\nu_2=0,1$ determine the spin structure.
For example, in the case $g_1(y)=g_-(2\pi N_\alpha y/L_y)$ and $g_2(x)=g_+(2\pi N_\beta x/L_x)$, the lift of these two transition functions over ${\rm Pin}_+$ group are
\begin{align}
\tilde{g}_1=\gamma_1\exp\left(\frac{\pi iN_\alpha y}{2L_y}[\sigma_1,\sigma_2]\right),\qquad
\tilde{g}_2=\exp\left(\frac{\pi iN_\beta x}{2L_x}[\sigma_1,\sigma_2]\right).
\end{align}
Here, $\sigma_1,\sigma_2$ are the Pauli matrices.
These lifted transition functions $\tilde{g}_1$ and $\tilde{g}_2$ satisfy the cocycle condition over a common relation $U_{11}\cap U_{31}\cap U_{13}$ if $N_\alpha-N_\beta=0$ mod $2$, and do not satisfy if $N_\alpha-N_\beta=1$ mod $2$.
If we confirm the cocycle conditions also in other common regions of the good covers, we find that the second Stiefel-Whitney class of torus whose $O(2)$-bundle satisfies $N_\alpha-N_\beta=0$ mod $2$ is the trivial element of $H^2(T^2,\mathbb{Z}_2)=\mathbb{Z}_2$, and the second Stiefel-Whitney class of a torus whose $O(2)$-bundle satisfies $N_\alpha-N_\beta=1$ mod $2$ are the generator of $H^2(T^2,\mathbb{Z}_2)$.

Now we consider the case $G=O(N\geq 2)$.
In the case $N=2n$, we assume that the $O(2n)$ transition functions along the two directions of the torus are given as follows:
\begin{align}
g_1(y)=&\left(
\begin{array}{ccc}
g_{\rho_1}\left(\frac{2\pi N_{\alpha_1}y}{L_y}\right)&&\\
&\ddots&\\
&&g_{\rho_n}\left(\frac{2\pi N_{\alpha_n}y}{L_y}\right)
\end{array}
\right),
\quad
g_2(x)=&\left(
\begin{array}{ccc}
g_{\kappa_1}\left(\frac{2\pi N_{\beta_1}x}{L_x}\right)&&\\
&\ddots&\\
&&g_{\kappa_n}\left(\frac{2\pi N_{\beta_n}x}{L_x}\right)
\end{array}
\right).\label{O(2k)=pf}
\end{align}
In the case $N=2n+1$, we assume that $O(2n+1)$ transition functions are
\begin{align}
g_1=&\left(
\begin{array}{cccc}
g_{\rho_1}\left(\frac{2\pi N_{\alpha_1}y}{L_y}\right)&&&\\
&\ddots&&\\
&&g_{\rho_n}\left(\frac{2\pi N_{\alpha_n}y}{L_y}\right)&\\
&&&1
\end{array}
\right),
\quad
g_2=&\left(
\begin{array}{cccc}
g_{\kappa_1}\left(\frac{2\pi N_{\beta_1}x}{L_x}\right)&&&\\
&\ddots&&\\
&&g_{\kappa_n}\left(\frac{2\pi N_{\beta_n}x}{L_x}\right)&\\
&&&1
\end{array}
\right).\label{O(2k+1)=pf}
\end{align}
We can verify that the second Stiefel-Whitney class of an $O(N)$-bundle over a torus whose transition functions are given by eq.(\ref{O(2k)=pf}) or eq.(\ref{O(2k+1)=pf}) is trivial if $\sum_j(N_{\alpha_j}-N_{\beta_j})=0$ mod $2$, and non-trivial if $\sum_j(N_{\alpha_j}-N_{\beta_j})=1$ mod $1$.
Since the Kronecker pairing of the second Stiefel-Whitney class and the fundamental class will be given in eq.(\ref{w2}), and the fact that the ${\rm Ker}(B{\rm det}_\ast)$ part of the bordism class in eq.(\ref{AHSS=2}) is spanned by tori,  the Kronecker pairing of the second Stiefel-Whitney class and the fundamental class is non-trivial on tori.
Therefore, the Kronecker pairing of the second Stiefel-Whitney class and the fundamental class on a torus is given by $(-1)^{\sum_j(N_{\alpha_j}-N_{\beta_j})}\in \mathbb{Z}_2$.

We obtained three independent bordism invariants $\eta[pt]\circ\Phi_{pt}$, ${\rm Arf}[\mathbb{Z}_2]\circ B{\rm det}_\ast$, and the second Stiefel-Whitney class $\left<w_2(E)\right>$.
We can confirm that these tori span all bordism classes $\Omega_2^{\rm Spin}(BO(N))=3\mathbb{Z}_2$ as listed in Table \ref{BO(N)-rep} .
\begin{table}[h]
  \caption{All bordism classes in $\Omega_2^{\rm Spin}(BO(N))$ spanned by tori}
  \label{BO(N)-rep}
  \centering
  \begin{tabular}{|c|c|c|c|c|}
     \hline
     $(\nu_1,\nu_2)$ & $({\rm det}g_1+\nu_1,{\rm det}g_2+\nu_2)$ & $\sum_j(N_{\alpha_j}-N_{\beta_j})$ &$(\eta[pt]\circ\Phi_{pt},\;{\rm Arf}[\mathbb{Z}_2]\circ B{\rm det}_\ast,\;\left<w_2(E)\right>)$\\
     \hline\hline
     $(0,0)$ & $(0,0)\:{\rm mod}\:2$ & $0\:{\rm mod}\:2$ & $(1,1,0)$\\
     \hline
      $(0,0)$ & $(0,0)\:{\rm mod}\:2$ & $1\:{\rm mod}\:2$ & $(1,1,1)$\\
     \hline
      $(0,0)$ & not $(0,0)\:{\rm mod}\:2$ & $0\:{\rm mod}\:2$ & $(1,0,0)$\\
     \hline
      $(0,0)$ & not $(0,0)\:{\rm mod}\:2$ & $1\:{\rm mod}\:2$ & $(1,0,1)$\\
     \hline%
      not $(0,0)$ & $(0,0)\:{\rm mod}\:2$ & $0\:{\rm mod}\:2$ & $(0,1,0)$\\
     \hline
     not $(0,0)$ & $(0,0)\:{\rm mod}\:2$ & $1\:{\rm mod}\:2$ & $(0,1,1)$\\
     \hline
      not $(0,0)$ & not $(0,0)\:{\rm mod}\:2$ & $0\:{\rm mod}\:2$ & $(0,0,0)$\\
     \hline
      not $(0,0)$ & not $(0,0)\:{\rm mod}\:2$ & $1\:{\rm mod}\:2$ & $(0,0,1)$\\
     \hline
         \end{tabular}
\end{table}

We will finally determine the global anomalies $\eta[O(N)]$ on these tori, which must be a linear combination of these three bordism invariants $\eta[pt]\circ\Phi_{pt}$, ${\rm Arf}[\mathbb{Z}_2]\circ B{\rm det}_\ast$, and the second Stiefel-Whitney class $\left<w_2(E)\right>$.

\subsection{$\eta$-Invariant of $O(N)$-bundles}
We will calculate $\eta[O(N)]$ in the case $N=2$ on a torus $x\sim x+L_x$, $y\sim y+L_y$.
The $O(2)$ gauge field can be written as
\begin{align}
A(x,y)=a(x,y)\left(
\begin{array}{cc}
0&1\\
-1&0
\end{array}
\right),\qquad
a(x,y)=a_x(x,y)dx+a_y(x,y)dy.\label{a(x,y)}
\end{align}
Because $\eta[O(2)]$ is a bordism invariant, this value does not depend on a specific choice of a gauge field.
We will calculate the number of zero-modes by fixing a specific gauge field on a torus.
We do not consider $(\rho,\kappa)=(+,+)$ case in eq.(\ref{gpm=2}) in this subsection, because we already studied the number of zero-modes in this case in section $4$.

We write the spinor components of a fermion as $\Psi=(\psi,\chi)$.
We also introduce the $O(2)$ components of a fermion as $\psi=e^{i\theta_1}(\psi_1,\psi_2)$ and $\chi=e^{i\theta_2}(\chi_1,\chi_2)$, and define
$\psi_\pm=e^{i\theta_1}(\psi_1\pm i\psi_2)$ and $\chi_\pm=e^{i\theta_2}(\chi_1\pm i\chi_2)$, where $e^{i\theta_1},e^{i\theta_2}\in U(1)$, and $\psi_1,\psi_2,\chi_1,\chi_2$ are valued in $\mathbb{R}$.
Then, the boundary conditions eq.(\ref{tf-1}) are written as
\begin{align}
\psi_\pm(L_x,y)=&(-1)^{\nu_1+\rho}e^{\mp\rho 2\pi iN_\alpha y/L_y}\psi_{\pm\rho}(0,y),
\qquad
\psi_\pm(x,L_y)=(-1)^{\nu_2+\kappa}e^{\mp\kappa 2\pi iN_\beta x/L_x}\psi_{\pm\kappa}(x,0),
\notag \\
\chi_\pm(L_x,y)=&(-1)^{\nu_1+\rho}e^{\mp\rho 2\pi iN_\alpha y/L_y}\chi_{\pm\rho}(0,y),
\qquad
\chi_\pm(x,L_y)=(-1)^{\nu_2+\kappa}e^{\mp\kappa 2\pi iN_\beta x/L_x}\chi_{\pm\kappa}(x,0),
\label{bc=4}
\end{align}
where $\rho,\kappa=\pm$ is defined in eq.(\ref{gpm=2}).
The zero-mode equation $\slashb{D}\Psi=0$ can be written as
\begin{align}
\left\{(\partial_x-i\partial_y)\mp ia_-\right\}\psi_\pm=0,\qquad
\left\{(\partial_x+i\partial_y)\mp ia_+\right\}\chi_\pm=0,
\label{zero=2}
\end{align}
Here, we define $a_\pm(x,y):=a_x(x,y)\pm ia_y(x,y)$, where $a_x(x,y)$ and $a_y(x,y)$ are defined in eq.(\ref{a(x,y)}).
In the case $(\rho,\kappa)=(-,+)$ snd $N_\beta=0$ are satisfied, we choose a gauge field as
\begin{align}
a_-(x,y)=&i\frac{2\pi N_\alpha}{L_y}\frac{x}{L_x}.
\label{gauge=10=2}
\end{align}
However, we cannot define a continuous gauge field $a_-(x,y)$ on local coordinates in other cases.
We will discuss the number of the solutions of the boundary conditions (\ref{bc=4}) and zero-mode equation (\ref{zero=2}) in the case $(\rho,\kappa)=(-,+)$ and $N_\beta=0$ are satisfied in Appendix B.
In the same way, we can consider the case $G=O(N\geq 2)$ where the transition functions along the two directions are given by eq.(\ref{O(2k)=pf}) or eq.(\ref{O(2k+1)=pf}).
In these cases, the zero-mode spaces are spanned by $n$ $O(2)$ zero-mode spaces.
Then, $\eta[O(2)]$ on these tori are summarized as Table \ref{eta=O(2)}.
\begin{table}[h]
 \caption{The $O(2)$ $\eta$-invariant on Tori}
 \label{eta=O(2)}
  \centering
  \begin{tabular}{|c|c|}
     \hline
     Tori&$\eta[O(2)]$\\
     \hline\hline
     Any $(\nu_1,\nu_2)$, $g_1(y)=g_+(2\pi N_\alpha y/L_y)$ and $g_2(x)=g_+(2\pi N_\beta x/L_x)$ & $N_\alpha+N_\beta$ mod $2$\\\hline
          $(\nu_1,\nu_2)=(0,0)$, $g_1(y)=g_-(2\pi N_\alpha y/L_y)$ and $g_2(x)=1$ & $N_\alpha+1$ mod $2$\\\hline
             $(\nu_1,\nu_2)=(1,0)$, $g_1(y)=g_-(2\pi N_\alpha y/L_y)$ and $g_2(x)=1$ & $N_\alpha+1$ mod $2$\\\hline
             $(\nu_1,\nu_2)=(0,1)$, $g_1(y)=g_-(2\pi N_\alpha y/L_y)$ and $g_2(x)=1$ & $N_\alpha$ mod $2$\\\hline
             $(\nu_1,\nu_2)=(1,1)$, $g_1(y)=g_-(2\pi N_\alpha y/L_y)$ and $g_2(x)=1$ & $N_\alpha$ mod $2$\\\hline
  \end{tabular}
\end{table}
By comparing with Table \ref{BO(N)-rep}, we find 
\begin{align}
\eta[O(2)]=\eta[pt]\circ\Phi_{pt}+{\rm Arf}[\mathbb{Z}_2]\circ B{\rm det}_\ast+\left<w_2(E)\right>.\label{etaO(2)=SUM}
\end{align}
We can generalize this result in the case $G=O(N)$:
\begin{align}
\eta[O(N)]=(N-1)\eta[pt]\circ\Phi_{pt}+{\rm Arf}[\mathbb{Z}_2]\circ B{\rm det}_\ast+\left<w_2(E)\right>.\label{etaO(N)=SUM}
\end{align}
Therefore, we obtained global anomalies corresponding to mapping tori whose $O(N)$ transition functions are given by eq.(\ref{O(2k)=pf}) or eq.(\ref{O(2k+1)=pf}).

\section{Conclusions and Discussions}
In this paper, we determined the range of global anomalies of a Majorana fermion in one dimension with symmetries $G=\mathbb{Z}_N$, $SO(N)$, and $O(N)$.
Based on the bordism group obtained in Appendix A, we constructed sufficient numbers of bordism invariants to obtain all bordism classes.

The results are summarized as follows.
In the case $G=\mathbb{Z}_N$, we confirmed that there are no global anomalies.
In the case $G=SO(N)$, we constructed and calculated bordism invariants $\eta[pt]\circ\Phi_{pt}$ and $\eta[SO(N)]$ in the case $N\geq 3$, and $c_1[U(1)]$ and $\eta[pt]\circ\Phi_{pt}$ in the case $N=2$, and determine the range of global anomalies by calculating $\eta[SO(N)]$ for all bordism classes.
In the case $G=SO(2)$, global anomalies are eliminated by adding the counter term defined in eq.(\ref{counter}).
In the case $G=O(N)$, we first found three bordism invariants $\eta[pt]\circ\Phi_{pt}$, ${\rm Arf}[\mathbb{Z}_2]\circ B{\rm det}_\ast$, and $\left<w_2(E)\right>$ to find all bordism classes, and then we obtained global anomalies correspond to those bordism classes.
We also confirmed that the range of the global anomalies determined by the anomaly inflow is equivalent to the range of global anomalies studied by acting symmetry transformations on the one-dimensional Hilbert space.

\section*{Acknowledgments}
The author thanks Yuji Tachikawa for suggesting the topic treated in this paper and advising her on many useful comments.
The author also thanks Kiyonori Gomi for advising her about the calculation of the bordism group and the construction of bordism invariants.
S.K is supported by Iwanami Fuujyukai Foundation.

\appendix

\section{Bordism Groups}
In Appendix A, we will calculate the bordism groups by using the Atiyah-Hirzebruch Spectral Sequence (AHSS).
We will first explain the Steenrod square, which is necessary to calculate the bordism groups by the AHSS.
Then, we will explain the method to calculate the bordism group by using the AHSS.
Finally, we will show the calculation of the bordism group in the cases $G=\mathbb{Z}_N$, $SO(N)$, and $O(N)$, and the unoriented bordism group in the case $G=O(N)$.

\subsection{Steenrod Square}
We first explain the Steenrod square based on p.$280$--p.$290$ in \cite{DK}, or \cite{Steen=1}.
For an arbitrary topological space $M$, we can define the $k$-th Steenrod Square $Sq^k:\:H^\ast(M,\mathbb{Z}_2)\to H^{\ast+k}(M,\mathbb{Z}_2)$, $k\in\mathbb{Z}_{\geq 0}$,
which satisfies the following conditions \cite{DK}:
\begin{description}
\item[a.]
$Sq^0(x)=x$,
\item[b.]
$Sq^i(x)=x^2$ if $x\in H^i(X;\mathbb{Z}_2)$,
\item[c.]
$Sq^i(x)=0$ if $x\in H^{i-p}(X;\mathbb{Z}_2),p>0$,
\item[d.]
$Sq^i(xy)=\sum_jSq^jxSq^{i-j}y$ (Cartan formula).
\end{description}
The Steenrod square satisfies the Wu formula \cite{Wu}:
\begin{align}
Sq^i(w_j)=\sum_{k=0}^i
\left(
\begin{array}{c}
(j-i)+(k-1)\\
k
\end{array}
\right)
w_{i-k}w_{j+k},
\label{Stiefel1}
\end{align}
where $w_i\in H^i(M,\mathbb{Z}_2)$ is the Stiefel-Whitney class.
The formula of the Steenrod square on general elements of $H^\ast(M,\mathbb{Z}_2)$ is given by the following relations \cite{Brown,Serre, Toda}:
\footnote{
We obtain an isomorphism $B\mathbb{Z}_p\times B\mathbb{Z}_q=B\mathbb{Z}_{pq}$ by the isomorphism $\mathbb{Z}_p\times \mathbb{Z}_q=\mathbb{Z}_{pq}$.
Then, by using the K\"{u}nneth formula, we find $H^\ast(B\mathbb{Z}_{2(2^h(2q+1))},\mathbb{Z}_2)=H^\ast(B\mathbb{Z}_{2^{h+1}},\mathbb{Z}_2)$. The $\mathbb{Z}_2$-coefficient cohomology $H^\ast(B\mathbb{Z}_{2^h},\mathbb{Z}_2)$ are written in p.205 of \cite{Serre} and p.79 of \cite{Toda}. }
\begin{align}
H^\ast(BO(n),\mathbb{Z}_2)=&\mathbb{Z}_2[w_1,\ldots,w_n],
\label{Hast-On}
\\
H^\ast(BSO(n),\mathbb{Z}_2)=&\mathbb{Z}_2[w_2,\ldots,w_n],\qquad w_1=0,
\label{Hast-SOn}
\\
H^\ast(B\mathbb{Z}_{2n},\mathbb{Z}_2)=&\mathbb{Z}_2[w,v]/(w^2-nv)
,\qquad
w\in H^1(B\mathbb{Z}_{2n},\mathbb{Z}_2),\qquad
v\in H^2(B\mathbb{Z}_{2n},\mathbb{Z}_2).
\label{Hast-2n}
\end{align}
For example, we find $H^\ast(B\mathbb{Z}_{2(2k+1)},\mathbb{Z}_2)=\mathbb{Z}_2[w]$.
By the Cartan formula, we obtain
\begin{align}
Sq^1(w^n)=w Sq^1(w^{n-1})+w^{n+1},\quad
Sq^2(w^n)=w^2Sq^1(w^{n-1})+w Sq^2(w^{n-1}).\label{2(2k+1)-app}
\end{align}
Here, $w\in H^1(B\mathbb{Z}_2,\mathbb{Z}_2)$.
From eq.(\ref{2(2k+1)-app}), we find
\begin{align}
Sq^1(w^n)=nw^{n+1},\qquad
Sq^2(w^n)=\frac{n(n-1)}{2}w^{n+2}.\label{Sq}
\end{align}
We can obtain the formula of the Steenrod square in the cases $H^\ast(B\mathbb{Z}_{4k},\mathbb{Z}_2)$, $H^\ast(BSO(n),\mathbb{Z}_2)$, and $H^\ast(BO(n),\mathbb{Z}_2)$ in a similar way.
Let us summarize these results.
In the case $G=\mathbb{Z}_{4k}$, the Steenrod square satisfies
\begin{align}
Sq^2(v^{k+1})=&vSq^2(v^k)+wvSq^1(v^k)+v^{k+2}
,\\
Sq^2(wv^{k+1})=&vSq^2(wv^k)+wvSq^1(wv^k)+wv^{k+2}.
\end{align}
In particular, $Sq^2:\:H^p(B\mathbb{Z}_{2n},\mathbb{Z}_2)\to H^{p+2}(B\mathbb{Z}_{2n},\mathbb{Z}_2)$ is zero if $p=0,1,4$.
In the cases $p=2,3$, $Sq^2:\:H^p(B\mathbb{Z}_{2n},\mathbb{Z}_2)\to H^{p+2}(B\mathbb{Z}_{2n},\mathbb{Z}_2)$ is an isomorphism.
In the case $G=SO(N)$, we find
\begin{align}
Sq^1(w_j)=(j-1)w_{j+1},\qquad
Sq^2(w_j)=
w_{2}w_{j},\qquad
j=2,\ldots,n.
\label{Sq2-SO(N)}
\end{align}
From the definition of $Sq^2$, $Sq^2:\:H^{p=0,1}(BSO(N),\mathbb{Z}_2)\to H^{p+2=2,3}(BSO(N),\mathbb{Z}_2)$ are zero map.
We also find that $Sq^2:\:H^2(BSO(N),\mathbb{Z}_2)\to H^{4}(BSO(N),\mathbb{Z}_2)$ is injective,
 $Sq^2:\:H^k(BSO(2),\mathbb{Z}_2)\to H^{k+2}(BSO(2),\mathbb{Z}_2)$ is a zero map if $k\geq 3$,
and $Sq^2:\:H^3(BSO(N\geq 3),\mathbb{Z}_2)\to H^{5}(BSO(N\geq 3),\mathbb{Z}_2)$ and $Sq^2:\:H^4(BSO(N\geq 3),\mathbb{Z}_2)\to H^{6}(BSO(N\geq 3),\mathbb{Z}_2)$ are injective.
In the case $G=O(N)$, $Sq^2:\:H^{p=0,1}(BO(N),\mathbb{Z}_2)\to H^{p+2=2,3}(BO(N),\mathbb{Z}_2)$ are zero maps, and $Sq^2:\:H^2(BO(N),\mathbb{Z}_2)\to H^{4}(BO(N),\mathbb{Z}_2)$ is injective.

\subsection{Atiyah-Hirzebruch Spectral Sequence (AHSS)}
We now explain the method to calculate the bordism group by using the Atiyah-Hirzebruch Spectral Sequence (AHSS) \cite{DK}.
Let us consider a Serre fibration $\{pt\}\to BG\to BG$, where $G$ is a group.
 By using theorem $9.6$ and section $9.3$ in \cite{DK}, there exists a spectral sequence $\{E^r_{p,q},d^r\}_{r\in\mathbb{Z}_{\geq 0};p,q\in\mathbb{Z}}$ and a filtration of $\Omega_d^{\rm Spin}(BG)$ which satisfy the following conditions:
 \begin{itemize}
 \item
 $E^2$ page of the AHSS is given as
$E^2_{p,q}=H_p(BG,\Omega_q^{\rm Spin}(pt))$.
\item
Filtration of $\Omega_d^{\rm Spin}(BG)$ is given as
\begin{align}
0=&F_{-1}\Omega_m^{\rm Spin}(BG)\subset F_0\Omega_m^{\rm Spin}(BG)\subset \ldots\subset F_m\Omega_m^{\rm Spin}(BG)=\Omega_m^{\rm Spin}(BG),
\label{AHSS}
\\
E^\infty_{k,n-k}=&\frac{F_k\Omega_n^{\rm Spin}(BG)}{F_{k-1}\Omega_n^{\rm Spin}(BG)}.
\label{Serre2}
\end{align}
\end{itemize}
Here, $\{E^r_{p,q},d^r\}_{r\in\mathbb{Z}_{\geq 0};p,q\in\mathbb{Z}}$ are chain complexes and $d^r:E^r_{p,q}\to E^r_{p-r,q+r-1}$ is the differential.
By the definition of a spectral sequence, $E^{r\geq 3}$ are given as
\begin{align}
E^{r+1}_{p,q}=H(E^r_{p,q},d^r)=\frac{{\rm ker}\;d^r:\:E^r_{p,q}\to E^r_{p-r,q+r-1}}{{\rm Im}\;d^r:\:E^r_{p+r,q-r+1}\to E^r_{p,q}}.
\label{AHSS-1}
\end{align}
By using eq.(\ref{AHSS}) and eq.(\ref{Serre2}), the bordism group $\Omega_d^{\rm Spin}(BG)$ is determined by $E^\infty$ as follows:
\begin{align}
\Omega_d^{\rm Spin}(BG)=&e(F_{d-1}\Omega_d^{\rm Spin}(BG),E^\infty_{d,0})
=e(e(F_{d-2}\Omega_d^{\rm Spin}(BG),E^\infty_{d-1,1}),E^\infty_{d,0})
\notag \\
=&e(e(\ldots (e(E^\infty_{0,d},E^\infty_{1,d-1}),E^\infty_{2,d-2}),\ldots ),E^\infty_{d-1,1}),E^\infty_{d,0}).
\label{E-infty}
\end{align}
Here, we denote by $e(A,B)$ an extension of $B$ by $A$:
\begin{align}
0\to A\to e(A,B)\to B\to 0.\label{expansion}
\end{align}

To determine the $E^\infty$ page, we should determine $d^2$.
For a topological space $X$, the universal coefficient theorem provides an isomorphism:
\begin{align}
H_p(X;\mathbb{Z}_2)\simeq {\rm Hom}_{\mathbb{Z}_2}(H^p(X;\mathbb{Z}_2),\mathbb{Z}_2).
\end{align}
Thus, we can introduce the following natural homomorphism as a dual of $Sq^2:H^p(X;\mathbb{Z}_2)\to H^{p+2}(X;\mathbb{Z}_2)$:
\begin{align}
(Sq^2)^\ast:\:
H_p(X;\mathbb{Z}_2)={\rm Hom}_{\mathbb{Z}_2}(H^p(X;\mathbb{Z}_2),\mathbb{Z}_2)
\to
H_{p-2}(X;\mathbb{Z}_2)={\rm Hom}_{\mathbb{Z}_2}(H^{p-2}(X;\mathbb{Z}_2),\mathbb{Z}_2).
\label{dual-Sq2}
\end{align}
We call $(Sq^2)^\ast$ dual of $Sq^2$.
We also introduce the ``reduction mod $2$" as the following natural isomorphism:
\begin{align}
r_2:=f\circ g;\qquad
H_p(X,\mathbb{Z})\xrightarrow{g}H_p(X;\mathbb{Z})\otimes \mathbb{Z}_2
\xrightarrow{f}H_p(X;\mathbb{Z}_2).
\label{r2}
\end{align}
Here, the map $g$ is a natural homomorphism, and the map $f$ appears in the universal coefficient theorem:
\begin{align}
1\to H_2(X;\mathbb{Z})\otimes\mathbb{Z}_2
\xrightarrow{f}
H_2(X;\mathbb{Z}_2)
\to{\rm Tor}(H_1(X;\mathbb{Z});\mathbb{Z}_2)\to 1.
\end{align}
Then, the differential $d^2$ is given as follows (See \cite{Steen=3}, p.$27$, Lemma $2.3.2$):
\begin{description}
\item[L1.]
$d^2:H_p(X;\Omega_1^{\rm Spin}(pt))\to H_{p-2}(X,\Omega_2^{\rm Spin}(pt))$ is equivalent to dual of 
$Sq^2:H^{p-2}(X,\mathbb{Z}_2)\to H^p(X,\mathbb{Z}_2)$.
\item[L2.]
$d^2:E^2_{p.0}=H_p(X;\Omega_0^{\rm Spin}(pt))\to E^2_{p-2,1}=H_{p-2}(X,\Omega_1^{\rm Spin}(pt))$
is equivalent to
\begin{align}
H_p(X;\mathbb{Z})\xrightarrow{r_2}
H_p(X;\mathbb{Z}_2)
\xrightarrow{(Sq^2)^\ast}
H_{p-2}(X;\mathbb{Z}_2).
\label{red2-Sq2}
\end{align}
\end{description}
We will calculate the bordism groups by determining $d^2$ in this way.

\subsection{Calculation of Bordism Groups $\Omega_d^{\rm Spin}(BG)$}
We will calculate the bordism groups in the cases $G=\mathbb{Z}_N$, $SO(N)$, and $O(N)$, by using the method of AHSS.
In this paper, we only calculate the bordism group in low dimensions.
Higher-dimensional examples are calculated in \cite{Garcia}.

\subsubsection*{The case $G=\mathbb{Z}_n$}
To obtain $\Omega_d^{\rm Spin}(B\mathbb{Z}_n)$,
we should first calculate $H_p(B\mathbb{Z}_n,\Omega_q^{\rm Spin}(pt))$ to use the AHSS method.
The universal coefficient theorem (See \cite{DK}, section $2.6$) provides the following exact sequence:
\begin{align}
0\to H_p(B\mathbb{Z}_n,\mathbb{Z})\otimes \Omega_q^{{\rm spin}}(pt)\to
H_p(B\mathbb{Z}_n;\Omega_q^{{\rm spin}}(pt))\to {\rm Tor}(H_{p-1}(B\mathbb{Z}_n,\mathbb{Z}),\Omega_q^{{\rm spin}}(pt))\to 0.\label{uni-coe}
\end{align}
Here, the torsion satisfies ${\rm Tor}(\mathbb{Z}_n,\mathbb{Z})=0$ and ${\rm Tor}(\mathbb{Z}_n,\mathbb{Z}_k)=\mathbb{Z}_n\otimes\mathbb{Z}_k=\mathbb{Z}_{{\rm gcd}(k,n)}$.
The point bordism group $\Omega_d^{\rm Spin}(pt)$ \cite{ABP67}  and the $\mathbb{Z}$-coefficient homology groups are given in Appendix C in \cite{Garcia}:
\begin{align}
\begin{array}{c|cccccccc}
q&0&1&2&3&4\\\hline
\Omega_q^{\rm Spin}(pt)&\mathbb{Z}&\mathbb{Z}_2&\mathbb{Z}_2&0&\mathbb{Z}
\end{array}
\qquad
H_p(B\mathbb{Z}_n,\mathbb{Z})=
\left\{
\begin{array}{ccc}
\mathbb{Z}&\qquad&p=0,\\
\mathbb{Z}_n&& p\in 2\mathbb{Z}+1,\\
0&&{\rm otherwise}.
\end{array}
\right.
\end{align}
Then, the $E^2$ pages is given as follows:
\begin{align}
\begin{array}{c|ccccccccc}
3&0&0&0&0&0\\
2&\mathbb{Z}_2&0&0&0&0\\
1&\mathbb{Z}_2&0&0&0&0\\
q=0&\mathbb{Z}&\mathbb{Z}_{2n+1}&0&\mathbb{Z}_{2n+1}&0\\\hline
E^2_{p,q}[\mathbb{Z}_{2n+1}]&p=0&1&2&3&4
\end{array}
\qquad
\begin{array}{c|ccccccccc}
3&0&0&0&0&0\\
2&\mathbb{Z}_2&\mathbb{Z}_2&\mathbb{Z}_2&\mathbb{Z}_2&\mathbb{Z}_2\\
1&\mathbb{Z}_2&\mathbb{Z}_2&\mathbb{Z}_2&\mathbb{Z}_2&\mathbb{Z}_2\\
q=0&\mathbb{Z}&\mathbb{Z}_{2n}&0&\mathbb{Z}_{2n}&0\\\hline
E^2_{p,q}[\mathbb{Z}_{2n}]&p=0&1&2&3&4
\end{array}
\label{uni-coe2}
\end{align}
where we denote by $E^2_{p,q}[G]$ the $E^2$ page of the AHSS of the bordism group $\Omega_d^{\rm Spin}(BG)$.

We first consider the case $G=\mathbb{Z}_{2n+1}$.
By using eq.(\ref{AHSS-1}), we clearly obtain $E^\infty_{p,q}=E^3_{p,q}=E^2_{p,q}$.
$\Omega_2^{\rm Spin}(B\mathbb{Z}_{2n+1})$ is given by substituting $E^\infty_{p,q}$ into eq.(\ref{E-infty}):
\begin{align}
\begin{array}{c|cccccccc}
q&0&1&2&3\\\hline
\Omega_q^{\rm Spin}(B\mathbb{Z}_{2n+1})&\mathbb{Z}&\mathbb{Z}_{2}\oplus\mathbb{Z}_{2n+1}&\mathbb{Z}_2&\mathbb{Z}_{2n+1}
\end{array}
\end{align}

In the case $G=\mathbb{Z}_{2n}$,
by using Lemma L.1 and Lemma L.2, we find $d_2:E^2_{p,1}\to E^2_{p-2,2}$, $p=2,3,6$ and $d_2:E^2_{p,0}\to E^2_{p-2,1}$, $p=2,3,4,6$ are zero maps,
$d_2:E^2_{p,1}\to E^2_{p-2,2}$, $p=4,5$ are isomorphisms,
and $d_2:E^2_{5,0}\to E^2_{3,1}$ are surjective.
Then, the $E^3$ page is determined as follows:
\begin{align}
\begin{array}{c|ccccccccc}
3&0&0&0&0&0
\\
2&\mathbb{Z}_2&\mathbb{Z}_2&0&0&\mathbb{Z}_2
\\
1&\mathbb{Z}_2&\mathbb{Z}_2&\mathbb{Z}_2&0&0
\\
q=0&\mathbb{Z}&\mathbb{Z}_{2n}&0&\mathbb{Z}_{2n}&0\\\hline
E_{p,q}^3&p=0&1&2&3&4
\end{array}
\qquad
\begin{array}{c|ccccccccc}
3&0&0&0&0&0
\\
2&\mathbb{Z}_2&\mathbb{Z}_2&0&0&\mathbb{Z}_2
\\
1&\mathbb{Z}_2&\mathbb{Z}_2&\mathbb{Z}_2&0&0
\\
q=0&\mathbb{Z}&\mathbb{Z}_{2n}&0&\mathbb{Z}_{2n}&0\\\hline
E_{p,q}^\infty&p=0&1&2&3&4
\label{E3-Z2n}
\end{array}
\end{align}
In the $E^3$ page, $d_3:\:E^3_{3,0}\to E^3_{0,2}$ is trivial.
By theorem $9.10$ in \cite{DK}, $E^\infty_{0,q}$ are given as
\begin{align}
E^\infty_{0,q}=E^2_{0,q}=H_0(B\mathbb{Z}_{2n},\Omega_q^{\rm Spin}(pt))=\Omega_q^{\rm Spin}(pt).
\end{align}
Therefore, the bordism group is given as follows:
\begin{align}
\begin{array}{c|cccccccc}
q&0&1&2&3\\\hline
\Omega_q^{\rm Spin}(B\mathbb{Z}_{2n})&\mathbb{Z}&\mathbb{Z}_{2}\oplus\mathbb{Z}_{2n}&\mathbb{Z}_2\oplus\mathbb{Z}_2&e(e(\mathbb{Z}_2,\mathbb{Z}_2),\mathbb{Z}_{2n})
\end{array}
\end{align}

\subsubsection*{The case $G=U(1)$}
By using $H_{2k}(BU(1),\mathbb{Z})=\mathbb{Z}$ and $H_{2k-1}(BU(1),\mathbb{Z})=0$ ($k\in\mathbb{Z}_{\geq 0}$) (See \cite{Garcia}, p.$21$),
and using the universal coefficient theorem, we obtain the $\mathbb{Z}_2$-coefficient homology.
Then, the $E^2$ page of $\Omega_d^{\rm Spin}(BU(1))$ is as follows:
\begin{align}
\begin{array}{c|ccccccc}
3&0&0&0&0&0&0&0
\\
2&\mathbb{Z}_2&0&\mathbb{Z}_2&0&\mathbb{Z}_2&0&\mathbb{Z}_2
\\
1&\mathbb{Z}_2&0&\mathbb{Z}_2&0&\mathbb{Z}_2&0&\mathbb{Z}_2
\\
q=0&\mathbb{Z}&0&\mathbb{Z}&0&\mathbb{Z}&0&\mathbb{Z}
\\\hline
E^2_{p,q}&p=0&1&2&3&4&5&6
\end{array}
\qquad
\begin{array}{c|ccccccc}
3&0&0&0&0&0&0
\\
2&\mathbb{Z}_2&0&0&0&\mathbb{Z}_2&0
\\
1&\mathbb{Z}_2&0&0&0&0&0
\\
q=0&\mathbb{Z}&0&\mathbb{Z}&0&\mathbb{Z}&0
\\\hline
E^3_{p,q}&p=0&1&2&3&4&5&6
\end{array}
\label{uni=U(1)}
\end{align}
By using Lemma L.1, we find that $d^2:E^2_{p,1}\to E^2_{p-2,2}$ are zero maps for $p=2,3,5,6$, and isomorphism for $p=4$.
In the same way, we obtain that $d^2:E^2_{4,0}\to E^2_{2,1}$ is surjective, and $d^2:E^2_{p,1}\to E^2_{p-2,2}$ are zero maps if $p=2,3,5,6$.
Then, the $E^3$ page is calculated as eq.(\ref{uni=U(1)}).
For $0\leq p,q\leq 3$, we find $E^\infty_{p,q}=E^3_{p,q}$.
Thus, the bordism group is as follows:
\begin{align}
\begin{array}{c|cccc}
q&0&1&2&3\\\hline
\Omega_q^{\rm Spin}(BU(1))&
\mathbb{Z}&\mathbb{Z}_2&\mathbb{Z}_2\oplus\mathbb{Z}&0
\end{array}
\label{Bor=U(1)}
\end{align}

\subsubsection*{The case $G=SO(N\geq 3)$}
In the case $G=SO(N\geq 3)$, the $\mathbb{Z}$-coefficient cohomology is given in \cite{Brown,Fes83}.
We obtain the $E^2$ page of $\Omega_d^{\rm Spin}(BSO(N\geq 3))$ by substituting $\mathbb{Z}$-coefficient cohomology into the universal coefficient theorem:
\begin{align}
\begin{array}{c|cccccc}
3&0&0&0&0&0&0
\\
2&\mathbb{Z}_2&0&\mathbb{Z}_2&\mathbb{Z}_2&\mathbb{Z}_2&\mathbb{Z}_2
\\
1&\mathbb{Z}_2&0&\mathbb{Z}_2&\mathbb{Z}_2&\mathbb{Z}_2&\mathbb{Z}_2
\\
q=0&\mathbb{Z}&0&\mathbb{Z}_2&0&\mathbb{Z}&\mathbb{Z}_2
\\\hline
E^2_{p,q}[N=3]&p=0&1&2&3&4&5
\end{array}
\qquad
\begin{array}{c|cccccc}
3&0&0&0&0&0&0
\\
2&\mathbb{Z}_2&0&\mathbb{Z}_2&\mathbb{Z}_2&\mathbb{Z}_2\oplus\mathbb{Z}_2&\mathbb{Z}_2
\\
1&\mathbb{Z}_2&0&\mathbb{Z}_2&\mathbb{Z}_2&\mathbb{Z}_2\oplus\mathbb{Z}_2&\mathbb{Z}_2
\\
q=0&\mathbb{Z}&0&\mathbb{Z}_2&0&\mathbb{Z}\oplus\mathbb{Z}&\mathbb{Z}_2
\\\hline
E^2_{p,q}[N=4]&p=0&1&2&3&4&5
\end{array}
\notag 
\end{align}
\begin{align}
\begin{array}{c|cccccc}
3&0&0&0&0&0&0
\\
2&\mathbb{Z}_2&0&\mathbb{Z}_2&\mathbb{Z}_2&\mathbb{Z}_2\oplus\mathbb{Z}_2&\mathbb{Z}_2\oplus\mathbb{Z}_2
\\
1&\mathbb{Z}_2&0&\mathbb{Z}_2&\mathbb{Z}_2&\mathbb{Z}_2\oplus\mathbb{Z}_2&\mathbb{Z}_2\oplus\mathbb{Z}_2
\\
q=0&\mathbb{Z}&0&\mathbb{Z}_2&0&\mathbb{Z}\oplus\mathbb{Z}_2&\mathbb{Z}_2
\\\hline
E^2_{p,q}[N\geq 5]&p=0&1&2&3&4&5
\end{array}
\label{uni=SO(N)}
\end{align}
Here, we write $E^2_{p,q}[N]:=H_p(BSO(N\geq 3),\Omega_q^{\rm Spin}(pt))$.
In the $E^2$ page,
$d^2:\:E^2_{2,0}\to E^2_{0,1}$ and $d^2:\:E^2_{3,0}\to E^2_{1,1}$ are zero maps, 
and $d^2:\:E^2_{4,0}\to E^2_{2,1}$ is surjective.
We obtain the $E^3$ pages as follows:
\begin{align}
&\begin{array}{c|cccccc}
3&0&0&0&0
\\
2&\mathbb{Z}_2&0&0&0
\\
1&\mathbb{Z}_2&0&0&0
\\
q=0&\mathbb{Z}&0&\mathbb{Z}_2&0
\\\hline
E^3_{p,q}[N=3]&p=0&1&2&3
\end{array}
\qquad
\begin{array}{c|cccccc}
3&0&0&0&0
\\
2&\mathbb{Z}_2&0&0&0
\\
1&\mathbb{Z}_2&0&0&0
\\
q=0&\mathbb{Z}&0&\mathbb{Z}_2&0
\\\hline
E^3_{p,q}[N=4]&p=0&1&2&3
\end{array}
\notag \\
&\begin{array}{c|cccccc}
3&0&0&0&0
\\
2&\mathbb{Z}_2&0&0&\ast
\\
1&\mathbb{Z}_2&0&0&0
\\
q=0&\mathbb{Z}&0&\mathbb{Z}_2&0
\\\hline
E^3_{p,q}[N\geq 5]&p=0&1&2&3
\end{array}
\end{align}
We find $E^\infty_{p,q}=E^3_{p,q}$ if $0\leq p,q\leq 3$,.
Finally, we obtain the bordism group as follows:
\begin{align}
\begin{array}{c|cccc}
q&0&1&2&3\\\hline
\Omega_q^{\rm Spin}(BSO(N\geq 3))&
\mathbb{Z}&\mathbb{Z}_2&\mathbb{Z}_2\oplus\mathbb{Z}_2&0
\end{array}
\label{Bor=SO(N)}
\end{align}

\subsubsection*{The case $G=O(N\geq 2)$}
The $\mathbb{Z}$-coefficient cohomology $H^\ast(BO(N),\mathbb{Z})$ are given as the table below \cite{Brown}:
\begin{align}
\begin{array}{c|ccccccccc}
p&0&1&2&3&4&5&6\\\hline
H^p(BO(2),\mathbb{Z})&\mathbb{Z}&0&\mathbb{Z}_2&\mathbb{Z}_2&\mathbb{Z}&0&\mathbb{Z}_2
\\
H^p(BO(3),\mathbb{Z})&\mathbb{Z}&0&\mathbb{Z}_2&\mathbb{Z}_2&\mathbb{Z}\oplus\mathbb{Z}_2&0&\mathbb{Z}_2
\\
H^p(BO(4),\mathbb{Z})&\mathbb{Z}&0&\mathbb{Z}_2&\mathbb{Z}_2&\mathbb{Z}\oplus\mathbb{Z}_2&\mathbb{Z}_2&\mathbb{Z}_2
\\
H^p(BO(n\geq 5),\mathbb{Z})&\mathbb{Z}&0&\mathbb{Z}_2&\mathbb{Z}_2&\mathbb{Z}\oplus\mathbb{Z}_2&\mathbb{Z}_2&\mathbb{Z}_2\oplus\mathbb{Z}_2
\\
\end{array}
\end{align}
Using the universal coefficient theorem, the $E^2$ pages are determined as follows.
\begin{align}
\begin{array}{c|ccccccc}
3&0&0&0&0&0&0
\\
2&\mathbb{Z}_2&\mathbb{Z}_2&2\mathbb{Z}_2&\mathbb{Z}_2&0&\mathbb{Z}_2
\\
1&\mathbb{Z}_2&\mathbb{Z}_2&2\mathbb{Z}_2&\mathbb{Z}_2&0&\mathbb{Z}_2
\\
q=0&\mathbb{Z}&\mathbb{Z}_2&\mathbb{Z}_2&0&\mathbb{Z}&\mathbb{Z}_2
\\\hline
E^2_{p,q}[O(2)]&p=0&1&2&3&4&5
\end{array}
\qquad
\begin{array}{c|ccccccc}
3&0&0&0&0&0&0
\\
2&\mathbb{Z}_2&\mathbb{Z}_2&2\mathbb{Z}_2&2\mathbb{Z}_2&\mathbb{Z}_2&\mathbb{Z}_2
\\
1&\mathbb{Z}_2&\mathbb{Z}_2&2\mathbb{Z}_2&2\mathbb{Z}_2&\mathbb{Z}_2&\mathbb{Z}_2
\\
q=0&\mathbb{Z}&\mathbb{Z}_2&\mathbb{Z}_2&\mathbb{Z}_2&\mathbb{Z}&\mathbb{Z}_2
\\\hline
E^2_{p,q}[O(3)]&p=0&1&2&3&4&5
\end{array}
\notag 
\end{align}
\begin{align}
\begin{array}{c|ccccccc}
3&0&0&0&0&0&0
\\
2&\mathbb{Z}_2&\mathbb{Z}_2&2\mathbb{Z}_2&2\mathbb{Z}_2&2\mathbb{Z}_2&2\mathbb{Z}_2
\\
1&\mathbb{Z}_2&\mathbb{Z}_2&2\mathbb{Z}_2&2\mathbb{Z}_2&2\mathbb{Z}_2&2\mathbb{Z}_2
\\
q=0&\mathbb{Z}&\mathbb{Z}_2&\mathbb{Z}_2&\mathbb{Z}_2&\mathbb{Z}\oplus\mathbb{Z}_2&\mathbb{Z}_2
\\\hline
E^2_{p,q}[O(4)]&p=0&1&2&3&4&5
\end{array}\notag 
\end{align}
\begin{align}
\begin{array}{c|ccccccc}
3&0&0&0&0&0&0
\\
2&\mathbb{Z}_2&\mathbb{Z}_2&2\mathbb{Z}_2&2\mathbb{Z}_2&2\mathbb{Z}_2&3\mathbb{Z}_2
\\
1&\mathbb{Z}_2&\mathbb{Z}_2&2\mathbb{Z}_2&2\mathbb{Z}_2&2\mathbb{Z}_2&3\mathbb{Z}_2
\\
q=0&\mathbb{Z}&\mathbb{Z}_2&\mathbb{Z}_2&\mathbb{Z}_2&\mathbb{Z}\oplus\mathbb{Z}_2&2\mathbb{Z}_2
\\\hline
E^2_{p,q}[O(N\geq 5)]&p=0&1&2&3&4&5
\end{array}
\notag 
\end{align}
Then, we obtain the $E^3$ pages, and finally, the bordism groups are determined as follows:
\begin{align}
\begin{array}{c|ccccccc}
3&0&0&0&0
\\
2&\mathbb{Z}_2&\mathbb{Z}_2&2\mathbb{Z}_2&\ast
\\
1&\mathbb{Z}_2&\mathbb{Z}_2&2\mathbb{Z}_2&\ast
\\
q=0&\mathbb{Z}&\mathbb{Z}_2&\mathbb{Z}_2&0
\\\hline
E^3_{p,q}[O(2)]&p=0&1&2&3
\end{array}
\qquad
\begin{array}{c|ccccccc}
3&0&0&0&0
\\
2&\mathbb{Z}_2&\mathbb{Z}_2&\mathbb{Z}_2&\ast
\\
1&\mathbb{Z}_2&\mathbb{Z}_2&\mathbb{Z}_2&\ast
\\
q=0&\mathbb{Z}&\mathbb{Z}_2&\mathbb{Z}_2&\mathbb{Z}_2
\\\hline
E^3_{p,q}[O(3)]&p=0&1&2&3
\end{array}
\notag 
\end{align}

\begin{align}
\begin{array}{c|ccccccc}
3&0&0&0&0
\\
2&\mathbb{Z}_2&\mathbb{Z}_2&0&\ast
\\
1&\mathbb{Z}_2&\mathbb{Z}_2&0&\ast
\\
q=0&\mathbb{Z}&\mathbb{Z}_2&\mathbb{Z}_2&\mathbb{Z}_2
\\\hline
E^3_{p,q}[O(n\geq 4)]&p=0&1&2&3
\end{array}
\qquad
\begin{array}{c|cccccccc}
q&0&1&2\\\hline
\Omega_q^{\rm Spin}(BO(n))&\mathbb{Z}&\mathbb{Z}_2\oplus\mathbb{Z}_2&\mathbb{Z}_2\oplus e(\mathbb{Z}_2,\mathbb{Z}_2)
\end{array}
\label{Bor-O(n)}
\end{align}

\subsubsection*{The case $\Omega_2^O(BO(N))$}
Finally, we calculate the bordism group with the unoriented stable structure.
We denote by $\tilde{E}^2_{p,q}=H_p(BO(N),\Omega_q^O(pt))$ the $E^2$ page of the bordism group $\Omega_d^O(BO(N))$, to distinguish the stable structures.
The point bordism group is given as 
\begin{align}
\Omega_0^O(pt)=\mathbb{Z}_2,\qquad
\Omega_1^O(pt)=0,\qquad
\Omega_2^O(pt)=\mathbb{Z}_2,\qquad
\Omega_3^O(pt)=0.
\end{align}
Then, the $E^2$ page is given as follows:
\begin{align}
\begin{array}{c|ccccccc}
3&0&0&0&0
\\
2&\mathbb{Z}_2&\mathbb{Z}_2&2\mathbb{Z}_2&\ast
\\
1&0&0&0&0
\\
q=0&\mathbb{Z}_2&\mathbb{Z}_2&2\mathbb{Z}_2&\ast
\\\hline
\tilde{E}^2_{p,q}[O(N)]&p=0&1&2&3
\end{array}
\end{align}
By using eq.(\ref{AHSS-1}), we obtain the $E^\infty$ page as follows:
\begin{align}
\tilde{E}^\infty_{0,2}=&\tilde{E}^2_{0,2}=\Omega_2^O(pt)=\mathbb{Z}_2,\qquad
\tilde{E}^\infty_{1,1}=0,\qquad
\tilde{E}^\infty_{2,0}=E^2_{2,0}=2\mathbb{Z}_2.
\end{align}
Therefore, the bordism group is given as
\begin{align}
\Omega_2^O(BO(N))=\Omega_2^O(pt)\oplus\tilde{\Omega}_2^O(BO(N))=\tilde{E}^2_{0,2}\oplus\tilde{E}^2_{2,0}
=\tilde{E}^2_{0,2}\oplus H_2(BO(N),\mathbb{Z}_2).
\label{borO}
\end{align}

\section{Zero-modes on $O(2)$-Spin bundles}
We will calculate the number of zero-modes by solving the equations eq.(\ref{bc=4}) and eq.(\ref{zero=2}).
In the following, we only discuss the $\psi_\pm$ part.
We can calculate the $\chi_\pm$ part in the same way.

In the case $(\rho,\kappa)=(-,+)$ and $N_\beta=0$, we choose the gauge field as eq.(\ref{gauge=10=2}).
The zero-mode equation is given as
\begin{align}
\left\{(\partial_x-i\partial_y)\pm
\frac{2\pi N_\alpha}{L_y}\frac{x}{L_x}
\right\}\psi_\pm=0.\label{zero=g1=1}
\end{align}
We introduce $D_\pm(x,y)$ as
\begin{align}
\psi_\pm(x,y)=&
e^{i\pi\nu_2 y/L_y}
D_\pm(x,y).
\label{zero=g1=2}
\end{align}
Then, the boundary conditions eq.(\ref{bc=4}) become as follows:
\begin{align}
D_\pm(L_x,y)=(-1)^{\nu_1+1}e^{\pm 2\pi iN_\alpha y/L_y}D_\mp(0,y),
\qquad
D_\pm(x,L_y)=&D_\pm(x,0).\label{D=g1=2}
\end{align}
Because of the periodicity of $D_\pm$ along the $y$-direction, $D_\pm$ are expanded by the Fourier expansion:
\begin{align}
D_\pm(x,y)=\sum_{k\in\mathbb{Z}}b_\pm^k(x)e^{2\pi i ky/L_y}.\label{g11=bpm1}
\end{align}
We obtain the conditions of $b_\pm^k(x)$ by substituting this expansion into the first equation of eq.(\ref{D=g1=2}) and eq.(\ref{zero=g1=1}):
\begin{align}
\left(\partial_x+\frac{\pi (2k+\nu_2)}{L_y}\pm
\frac{2\pi N_\alpha}{L_y}\frac{x}{L_x}
\right)b_\pm^k(x)=0,\qquad
b_\pm^{k\pm N_\alpha}(L_x)=(-1)^{\nu_1+1}b_\mp^k(0).\label{g11=bpm2}
\end{align}
The solutions of eq.(\ref{g11=bpm2}) satisfy the following conditions:
\begin{align}
b_\pm^k(x)=&b_\pm^k\exp\left(-\frac{\pi(2k+\nu_2)}{L_y}x\mp\frac{\pi N_\alpha}{L_xL_y}x^2\right),\\
b_+^{k+N_\alpha}=&(-1)^{\nu_1+1}\exp\left(\frac{\pi(2k+N_\alpha+\nu_2)L_x}{L_y}\right)b_-^k,\label{101}
\\
b_-^{k-N_\alpha}=&(-1)^{\nu_1+1}\exp\left(\frac{\pi(2k-N_\alpha+\nu_2)L_x}{L_y}\right)b_+^k.\label{100}
\end{align}
If we substitute eq.(\ref{100}) into eq.(\ref{101}), or substitute eq.(\ref{101}) into eq.(\ref{100}), we find
\begin{align}
b_\pm^{k\pm N_\alpha}=\exp\left(\frac{2\pi(2k\pm N_\alpha+\nu_2)L_x}{L_y}\right)b_\pm^{k\pm N_\alpha}.\label{102}
\end{align}
Therefore, we find that $b_\pm^{k\pm N_\alpha}\neq 0$ if and only if $\exp\left(\frac{2\pi(2k\pm N_\alpha+\nu_2)L_x}{L_y}\right)=1$ is satisfied.
Since the argument of this phase is a real number, there are two solutions in the case $N_\alpha+\nu_2$ is an even integer, and otherwise, there are no solutions.

\section{Stiefel-Whitney class}
It is known that the Kronecker pairing of the Stiefel-Whitney class and the fundamental class is a bordism invariant \cite{Pnot=47}.
We will show that the first Stiefel-Whitney class is trivial,
and the second Stiefel-Whitney class is equivalent to $\eta[O(N)]$ for two dimensions.
\footnote{The proof of these statement are based on the idea of Kiyonori Gomi.}

We can construct a bordism invariant by using Kronecker pairing and the Stiefel-Whitney class \cite{Pnot=47}.
Introducing integers $m_1,\ldots,m_N\in\mathbb{Z}_{\geq 0}$ which satisfy $\sum_{j=1}^Njm_j={\rm dim}M$, we obtain a bordism invariant as follows:
\begin{align}
W_{m_1,\ldots,m_N}:\:\Omega_d^{\rm Spin}(BO(N))\to\mathbb{Z}_2,\qquad
[M,\phi_s,f_E]\to\left<w_1(E)^{m_1}\ldots w_N(E)^{m_N},[M]\right>.\label{Bor=W}
\end{align}
Here we denote by $[M]\in H^{{\rm dim}M}(M,\mathbb{Z})=\mathbb{Z}$ the fundamental class,
$w_k(E)\in H^k(M,\mathbb{Z}_2)$ is the $k$-th Stiefel-Whitney class of a vector bundle $E\to M$, and $\left<\quad,\quad\right>$ is the Kronecker pairing (See \cite{DK}, section $2.6$):
\begin{align}
\left<\quad,\quad\right>:\:
H^{{\rm dim}M}(M,\mathbb{Z}_2)\times H_{{\rm dim}M}(M,\mathbb{Z})
\to\mathbb{Z}_2.\label{paring=1}
\end{align}
The adjoint of eq.(\ref{paring=1}) is a map $H^{{\rm dim}M}(M,\mathbb{Z}_2)\to {\rm Hom}(H_{{\rm dim}M}(M,\mathbb{Z}),\mathbb{Z}_2)$ which satisfies the universal coefficient theorem:
\begin{align}
0\to {\rm Ext}(H_{{\rm dim}M-1}(M,\mathbb{Z}),\mathbb{Z}_2)
\to H^{{\rm dim}M}(M,\mathbb{Z}_2)\to {\rm Hom}(H_{{\rm dim}M}(M,\mathbb{Z}),\mathbb{Z}_2)
\to 0.\label{paring=2}
\end{align}
In particular, in the case ${\rm dim}M=2$, these bordism invariants are $\left<w_1(E)^2,[M]\right>$ and $\left<w_2(E),[M]\right>$.
Below, we assume $N\geq 2$.

\subsection{First Stiefel-Whitney class square: $\left<w_1(E)^2,[M]\right>=0$}
Let us consider the first Stiefel-Whitney class square.
It is enough to consider $\left<w_1(E)^2,[M]\right>$ on tori because tori span all bordism classes (In section $6$, we confirmed that tori span all bordism classes in Table \ref{BO(N)-rep} by using three bordism classes $\eta[pt]\circ\Phi_{pt},{\rm Arf}[\mathbb{Z}_2]\circ B{\rm det}_\ast$, and the second Stiefel-Whitney class. In Table \ref{BO(N)-rep}, we determine the second Stiefel-Whitney class by using Appendix C.2. In the proof of Appendix C.2, we will use the fact that the first Stiefel-Whitney class is trivial.
But we can also confirm that all bordism classes are spanned by tori by calculating $\eta[pt]\circ\Phi_{pt},{\rm Arf}[\mathbb{Z}_2]\circ B{\rm det}_\ast$, and $\eta[O(N)]$ on tori.).
By using the K\"{u}nneth formula, we find isomorphisms:
\begin{align}
H^1(S^1,\mathbb{Z}_2)\oplus H^1(S^1,\mathbb{Z}_2)\xrightarrow{\simeq \times} H^1(T^2,\mathbb{Z}_2),\quad
 H^1(S^1,\mathbb{Z}_2)\otimes H^1(S^1,\mathbb{Z}_2)\xrightarrow{\simeq \times}H^2(T^2,\mathbb{Z}_2).\label{Kunneth1}
\end{align}
The generator of $H^1(S^1,\mathbb{Z}_2)=\mathbb{Z}_2$ is $w_1(l)\in H^1(S^1,\mathbb{Z}_2)$.
Here, $l\to S^1$ is the non-trivial real line bundle.
From $w_1(l)^2=0$, we obtain
$H^\ast(S^1,\mathbb{Z}_2)=\mathbb{Z}_2[w_1(l)]/(w_1(l)^2)$.
By combining $H^\ast(S^1,\mathbb{Z}_2)=\mathbb{Z}_2[w_1(l)]/(w_1(l)^2)$ and eq.(\ref{Kunneth1}), the cohomology group $ H^\ast(T^2,\mathbb{Z}_2)$ can be written as a $\mathbb{Z}_2$-coefficient polynomial ring:
\begin{align}
 H^\ast(T^2,\mathbb{Z}_2)\simeq\mathbb{Z}_2[\alpha_1,\alpha_2]/(\alpha_1^2,\alpha_2^2).\label{Stiefel=1}
\end{align}
Here, we introduce $\alpha_1,\alpha_2\in H^1(T^2,\mathbb{Z}_2)$ as the first Stiefel-Whitney class of the pullback bundle $p_i^\ast(l)$ of the line bundle $l\to S^1$ by the projection $p_i:T^2\to S^1$, $i=1,2$.
By using eq.(\ref{Stiefel=1}), we can expand the first Stiefel-Whitney class of an $O(N)$-bundle on a torus $E\to T^2$
as $w_1(E)=q_1\alpha_1+q_2\alpha_2$, 
$q_1,q_2\in\mathbb{Z}_2$.
Then the square of $w_1(E)$ is trivial:
\begin{align}
w_1(E)^2=(q_1\alpha_1+q_2\alpha_2)^2=q_1^2\alpha_1^2+q_2^2\alpha_2^2+2q_1q_2\alpha_1\alpha_2=0\in H^2(T^2,\mathbb{Z}_2).
\end{align}
Because tori span all bordism classes, we find that $\left<w_1(E),[M]\right>$ is trivial for all two-dimensional manifold $M$.

\subsection{Second Stiefel-Whitney class}
We will claim that 
\begin{align}
\left<w_2(E),\ast\right>:\:
\Omega_2^{\rm Spin}(BO(N))=E^2_{0,2}\oplus E^2_{1,1}\oplus E^2_{2,0}\to\mathbb{Z}_2,\quad
(p_0,p_1,p_2)\to p_2,\quad
p_j\in\mathbb{Z}_2.\label{w2}
\end{align}
It is known that $O(1)=\mathbb{Z}_2$-bundle has a ${\rm Pin}^+$ structure (See \cite{pin}, Lemma $1.2$).
It is also known that the condition $w_2(E)=0$ for an $O(N)$-bundle $E$ is equivalent to the condition that the $O(N)$-bundle $E$ have a ${\rm Pin}^+$ structure (See \cite{pin}, Lemma $1.3$).
Then, by using the isomorphisms eq.(\ref{AHSS=2}), we can confirm that $E^2_{0,2}$ part and $E^2_{1,1}$ part of the bordism group in eq.(\ref{AHSS=direct}) do not contribute to $\left<w_2(E),[M]\right>$.

Because the definition of this bordism invariant $\left<w_2(E),[M]\right>$ does not depend on the spin structure of the bordism group, we can calculate eq.(\ref{w2}) by forgetting the spin structure.
Let us introduce the spin forgetful map and the orientation forgetful map:
\begin{align}
{\cal F}_{{\rm Spin}\to SO}&:H_p(BO(N),\Omega_q^{\rm Spin}(pt))\to H_p(BO(N),\Omega_q^{SO}(pt)),
\notag \\
{\cal F}_{SO\to O}&:H_p(BO(N),\Omega_q^{SO}(pt))\to H_p(BO(N),\Omega_q^{O}(pt)).\label{forget=1}
\end{align}
It is enough to consider the case $(p,q)=(2,0)$ in eq.(\ref{forget=1}).
By using $\Omega_0^{\rm Spin}(pt)=\Omega_0^{SO}(pt)=\mathbb{Z}$ and $\Omega_0^O(pt)=\mathbb{Z}_2$, we obtain that ${\cal F}_{{\rm Spin}\to SO}$ is an isomorphism, and ${\cal F}_{SO\to O}$ is the reduction mod $2$.
This is because the induced map of homology by a group homomorphism $\mathbb{Z}\to\mathbb{Z}_2$ is the 
reduction mod $2$, $r_2$.
By using $H_1(BO(N);\mathbb{Z})=\mathbb{Z}_2$, $r_2$ is injective.
Therefore, ${\cal F}_{SO\to O}\circ{\cal F}_{{\rm Spin}\to SO}$ becomes as follows:
\begin{align}
{\cal F}_{SO\to O}\circ{\cal F}_{{\rm Spin}\to SO}:\:
E^2_{2,0}=\mathbb{Z}_2\to& \tilde{E}^2_{2,0}=\mathbb{Z}_2\oplus\mathbb{Z}_2,\qquad
x\to(x,0)\:{\rm or}\:(0,x).\label{forget=(1)}
\end{align}
Here, we denote by $E^2_{p,q}$ and $\tilde{E}^2_{p,q}$ the $E^2$ pages of $\Omega_2^{\rm Spin}(BO(N))$ and $\Omega_2^O(BO(N))$.
Because $\tilde{E}^2_{0,2}=\Omega_2^O(pt)$ and $\tilde{E}^2_{1,1}=0$, we find that only $\tilde{E}^2_{2,0}$ part of the bordism group  $\Omega_2^O(BO(N))$ contributes to $\left<w_1^2(\phi^\ast(EO(N))),[M]\right>$ and $\left<w_2(\phi^\ast(EO(N))),[M]\right>$.

Now, we calculate $\left<w_1^2(\phi^\ast(EO(N))),[M]\right>$ and $\left<w_2(\phi^\ast(EO(N))),[M]\right>$ on $\Omega_2^O(BO(N))$.
We denote by $l\to\mathbb{R}P^2$ a non-trivial line bundle, and
consider the following two elements of $\Omega_2^O(BO(N))$:
\begin{align}
E:=l\oplus\underline{\mathbb{R}}^{N-1},\qquad
F:=l\oplus l\oplus\underline{\mathbb{R}}^{N-2}.
\end{align}
The $\mathbb{Z}_2$-coefficient cohomology on $\mathbb{R}P^2$ can be written as
\begin{align}
H^\ast(\mathbb{R}P^2,\mathbb{Z}_2)\simeq\mathbb{Z}_2[w_1(l)]/(w_1(l)^3).\label{RP2}
\end{align}
Then, the Stiefel-Whitney classes of these bundles are given as
\begin{align}
w_1(E)=w_1(l),\qquad
w_2(E)=0,\qquad
w_1(F)=0,\qquad
w_2(F)=w_1(l)^2.\label{line=2}
\end{align}
Since the Kronecker paring $\left<\quad,\quad\right>:\:
H^2(\mathbb{R}P^2,\mathbb{Z}_2)\times H_2(\mathbb{R}P^2,\mathbb{Z})\to \mathbb{Z}_2$ is the dual of eq.(\ref{paring=2}), this Kronecker paring is surjective.
We obtain
\begin{align}
\left<w_1(l)^2,[\mathbb{R}P^2]\right>=-1\in\{\pm 1\}=\mathbb{Z}_2
.\label{line=1}
\end{align}
By combining eq.(\ref{line=2}) and eq.(\ref{line=1}), we find
\begin{align}
\left<w_1(E)^2,[\mathbb{R}P^2]\right>=\left<w_2(F),[\mathbb{R}P^2]\right>=-1,\qquad
\left<w_2(E),[\mathbb{R}P^2]\right>=
\left<w_1(F)^2,[\mathbb{R}P^2]\right>=1.\label{w2=nont}
\end{align}
Therefore, $\left<w_1^2(\phi^\ast(EO(N))),[M]\right>$ and $\left<w_2(\phi^\ast(EO(N))),[M]\right>$ cause two independent injective maps on $\tilde{E}^2_{2,0}=\mathbb{Z}_2\oplus\mathbb{Z}_2\subset\Omega_2^O(BO(N))$.
By combining this result and eq.(\ref{forget=(1)}), the remaining problem is to decide the image of eq.(\ref{forget=(1)}).
Because $w_1(E)^2$ is trivial on $\Omega_2^{\rm Spin}(BO(N))$, the image of eq.(\ref{forget=(1)}) is the $\mathbb{Z}_2\subset \tilde{E}^2_{2,0}$, on which $\left<w_2(\phi^\ast(EO(N))),[M]\right>$ is non-trivial.
Thus, we conclude that $\left<w_2(\phi^\ast(EO(N))),[M]\right>$ is non-trivial on $\Omega_2^{\rm Spin}(BO(N))$, proving eq.(\ref{w2}).


\begin{thebibliography}{200}
\bibitem{Callan=Harvey}
C. G. Callan, Jr.  and J. A. Harvey,
Anomalies And Fermion Zero-Modes On Strings And Domain Walls,
Nucl. Phys. B250 (1985) 427-36.

\bibitem{Fad-Shat}
L. D. Faddeev, S. L. Shatashvili,
Algebraic and Hamiltonian Methods in the Theory of Nonabelian Anomalies, 
Theor. Math. Phys. 60 (1985) 770-778.

\bibitem{Jackiw=CS}
R. Jackiw, 
Topological Investigations Of Quantized Gauge Theories,
in B. S. DeWitt et. al., eds., Relativity, Groups, and Topology,II, Les Houches 1983, reprinted in updated form in S. B. Treiman et. al., eds., Current Algebra and Anomalies (World-Scientific, 1985).

\bibitem{Zumino}
B. Zumino, 
Chiral Anomalies In Differential Geometry,
in B. S. DeWitt et. al., eds., Relativity, Groups, and Topology,II, Les Houches 1983, reprinted in S. B. Treiman et. al., eds., Current Algebra and Anomalies (World-Scientific, 1985).

\bibitem{Stora}
R. Stora, 
Algebraic structure and topological origin of anomalies, in: Recent Progress in gauge theories, G. Lehman and al. Eds, New York, Plenum (1984).

\bibitem{Old-SU(2)}
E. Witten, 
An $SU(2)$ Anomaly,
Phys. Lett. 117B (1982) 324-8.

\bibitem{APS1}
M. F. Atiyah, V. K. Patodi and I. M. Singer, 
Spectral Asymmetry and Riemannian Geometry 1,
Math. Proc. Cambridge Phil. Soc. 77(1975)43.

\bibitem{APS2}
M. F. Atiyah, V. K. Patodi and I. M. Singer, 
Spectral Asymmetry and Riemannian Geometry 2,
Math. Proc. Cambridge Phil. Soc. 78(1976)405.

\bibitem{APS3}
M. F. Atiyah, V. K. Patodi and I. M. Singer, 
Spectral Asymmetry and Riemannian Geometry 3,
Math. Proc. Cambridge Phil. Soc. 79(1976)71-99.

\bibitem{GG=Witten}
E. Witten,
Global Gravitational Anomalies,
Commun. Math. Phys. 100, 197 (1985).

\bibitem{Dai-Freed}
X. Dai and D. S. Freed, 
Eta Invariants and Determinant Lines,
J. Math. Phys. 35, 5155 (1994),
arXiv:hep-th/9405012.

\bibitem{KY}
K. Yonekura,
Dai-Freed theorem and topological phases of matter,
JHEP 09 (2016) 022, 
arXiv:1607.01873 [hep-th].

\bibitem{EW}
E. Witten,
Fermion Path Integral and Topological Phases,
Rev. Mod. Phys. 88, 35001 (2016), 
arXiv:1508.04715 [cond-mat.mes-hall].

\bibitem{KY=EW}
E. Witten, K. Yonekura,
Anomaly Inflow and the $\eta$-Invariant,
arXiv:1909.08775 [hep-th].

\bibitem{Ati-TQFT=88}
M.Atiyah,
Topological quantum field theories,
Publications math\'{e}matiques de I'IH\'{E}S, tome 68 (1988), p.175-186.

\bibitem{Ryu-Moo=10}
S. Ryu, J. E. Moore, and A. W. W. Ludwig, Electromagnetic and gravitational responses and anomalies in topological insulators and superconductors, Phys. Rev. B85 (2012) 045104, arXiv:1010.0936 [cond-mat.str-el].

\bibitem{Wen=13}
X.-G. Wen, Classifying gauge anomalies through symmetry-protected trivial orders and classifying gravitational anomalies through topological orders, Phys. Rev. D88 (2013), no. 4 045013, arXiv:1303.1803 [hep-th].

\bibitem{Kapustin2}
A. Kapustin,
Symmetry Protected Topological Phases, Anomalies, and Cobordisms: Beyond Group Cohomology,
arXiv:1403.1467 [cond-mat.str-el].

\bibitem{Kapu-Tho=14}
A.Kapustin, R.Thorngren,
Anomalies of discrete symmetries in various dimensions and group cohomology,
arXiv:1404.3230 [hep-th].

\bibitem{WW=14}
J. C. Wang, Z.-C. Gu, and X.-G. Wen, Field theory representation of gauge-gravity symmetry-protected topological invariants, group cohomology and beyond, Phys. Rev. Lett. 114 (2015), no. 3 031601, arXiv:1405.7689 [cond-mat.str-el].

\bibitem{Kapustin}
A. Kapustin, R. Thorngren, A. Turzillo, Z. Wang,
Fermionic Symmetry Protected Topological Phases and Cobordisms,
JHEP 12 (2015) 052,
arXiv:1406.7329 [cond-mat.str-el].

\bibitem{CGS=15}
C.-T. Hsieh, G. Y. Cho, and S. Ryu, Global anomalies on the surface of fermionic symmetry-protected topological phases in $(3+1)$ dimensions, Phys. Rev. B93 (2016), no. 7 075135, arXiv:1503.01411 [cond-mat.str-el].

\bibitem{Wit=16}
E. Witten, The ``Parity” Anomaly On An Unorientable Manifold, Phys. Rev. B94 (2016), no. 19 195150, arXiv:1605.02391 [hep-th].

\bibitem{JWE-17}
J. Wang, X.-G. Wen, and E. Witten, Symmetric Gapped Interfaces of SPT and SET States: Systematic Constructions, Phys. Rev. X8 (2018) 031048,
arXiv:1705.06728 [cond-mat.str-el].

\bibitem{Tachikawa=17}
Y.Tachikawa,
On gauging finite subgroups,
SciPost Phys. 8, 015 (2020)
arXiv:1712.09542 [hep-th].

\bibitem{Fre-Hop=16}
D. S. Freed and M. J. Hopkins, Reflection positivity and invertible topological phases, 
Geom. Topol. 25 (2021) 1165-1330,
arXiv:1604.06527 [hep-th].

\bibitem{KY=18}
K. Yonekura,
On the cobordism classification of symmetry protected topological phases,
Commun.Math.Phys. 368 (2019) 3, 1121-1173,
arXiv:1803.10796 [hep-th].

\bibitem{FerSPT18}
M. Guo, K. Ohmori, P. Putrov, Z. Wan, and J. Wang, Fermionic Finite-Group Gauge Theories and Interacting Symmetric/Crystalline Orders via Cobordisms, arXiv:1812.11959 [hep-th].

\bibitem{RKY=19}
R. Kobayashi, K. Ohmori, Y. Tachikawa,
On gapped boundaries for SPT phases beyond group cohomology,
JHEP 11 (2019) 131, 
arXiv:1905.05391 [cond-mat.str-el].

\bibitem{Cor-Ohm=20}
C.C\'{o}rdova, K.Ohmori,
Anomaly Obstructions to Symmetry Preserving Gapped Phases,
arXiv:1910.04962 [hep-th].

\bibitem{discrete}
C.-T.Hsieh, 
Discrete Gauge Anomalies Revised,
arXiv:1808.02881 [hep-th].

\bibitem{Garcia}
I. Garc\'{i}a-Etxebarria, M. Montero,
Dai-Freed Anomalies in Particle Physics,
JHEP 08 (2019) 003,
arXiv:1808.00009 [hep-th].

\bibitem{Spherical}
P. B. Gilkey,
The Geometry of Spherical Space Form Groups.

\bibitem{AHSS}
M. F. Atiyah, F. Hirzebruch, 
Vector bundles and homogeneous spaces,
Matematica, 6:2 (1962), 3-39; Proc. Sympos. Pure Math.,
Vol. III, American Mathematical Society, Providence, R.I., 3 (1961), 7-38.

\bibitem{String=WS}
J. Kaidi, J. Parra-Martinez, Y. Tachikawa with mathematical Appendix by A. Debray,
Topological Superconductor on Superstring Worldsheets,
SciPost Phys. 9 (2020) 10,
arXiv:1911.11780 [hep-th].

\bibitem{Arf-1}
C. Arf, Untersuchungen \"{u}ber quadratische Formen in K\"{o}rpern der Charakteristik 2. 1,
J. Reine Angew. Math. 183 (1941) 148-167.

\bibitem{Arf-2}
D. Johnson, 
Spin structures and quadratic forms on surfaces,
Journal of the London Mathematical Society 2 (1980) 365-373.

\bibitem{Brown=72}
E. H. Brown, Jr., Generalizations of the Kervaire invariant,
Ann. of Math. (2) 95 (1972) 368-383.

\bibitem{ABK-1}
A. Turzillo, Diagrammatic State Sums for 2D Pin-Minus TQFTs, 
JHEP 03 (2020) 019,
arXiv:1811.12654 [math.QA].

\bibitem{ABK-2}
A. Debray and S. Gunningham, The Arf-Brown TQFT of ${\rm Pin}^-$ Surfaces, 
arXiv:1803.11183 [math-ph].

\bibitem{Kob=19}
R. Kobayashi, 
Pin TQFT and Grassmann Integral, 
arXiv:1905.05902 [cond-mat.str-el].

\bibitem{2d}
A. Grigoletto, P. Tutrov,
Spin-cobordisms, surgeries and fermionic modular bootstrap,
arXiv:2106.16247 [hep-th].

\bibitem{DK}
J. F. Davis, P. Kirk, 
Lecture Notes in Algebraic Topology, 
American Mathematical Society, Graduate Studies in Mathematics, Vol 35.

\bibitem{Hil=DGG21}
D.Delmastro, D.Gaiotto, J.Gomis,
Global Anomalies on the Hilbert Space,
arXiv:2101.02218 [hep-th].

\bibitem{AS=index}
M. F. Atiyah and I. M. Singer, The index of elliptic operators on compact manifolds, Bull. Amer. Math. Soc. 69 (1963), 422-433.

\bibitem{Steen=1}
A. Beaudry, J. A. Campbell,
A Guide for Computing Stable Homotopy Groups,
arXiv:1801.07530 [math.AT].

\bibitem{Wu}
J. Milnor and S. Stasheff, Characteristic Classes, No. 76 of Annals of Mathematics Studies, Princeton University Press, 1974.

\bibitem{Brown}
E. H. Brown, The cohomology of $BSO_n$ and $BO_n$ with integer coefficients,
Proceedings of the American Mathematical Society
Vol. 85, No. 2 (Jun., 1982), pp. 283-288.

\bibitem{Serre}
J.-P. Serre, 
Cohomologie modulo 2 des complexes d'Eilenberg-MacLane. Comment. Math. Helv. 27 (1953), 198-232.

\bibitem{Toda}
H. Toda,
Cohomology of classifying spaces,
Homotopy theory and related topics (Kyoto, 1984), 75-108,
Adv. Stud. Pure Math., 9, North-Holland, Amsterdam, 1987.

\bibitem{Steen=3}
P.Teichner,
TOPOLOGICAL FOUR-MANIFOLDS WITH FINITE FUNDAMENTAL GROUP,
Ph.D. Thesis,
\url{https://math.berkeley.edu/~teichner/Papers/phd.pdf}

\bibitem{ABP67}
D. W. Anderson, E. H. Brown, Jr., and F. P. Peterson, The structure of the Spin cobordism ring, 
Ann. of Math. (2) 86 (1967) 271-298.

\bibitem{Fes83}
M. Feshbach, 
The inregral cohomology ring of the classifying spaces of $O(n)$ and $SO(n)$,
Indiana Univ. Math. J. 32 (1983) 511-516.

\bibitem{Pnot=47}
L. S. Pontrjagin, 
Characteristic cycles on differentiable manifolds,
Math. Sbornik N. S. (1947) 21, (63) 233-284.

\bibitem{pin}
R. C. Kirby, L. R. Taylor,
Pin Structures on Low-dimensional Manifolds,
Cambridge University Press, pp.177-242.


\end{thebibliography}
\end{document}